Department of Computer Science

# Underpinnings of Digital-photo interaction in Computer-mediated platforms

Aqdas Amin Malik

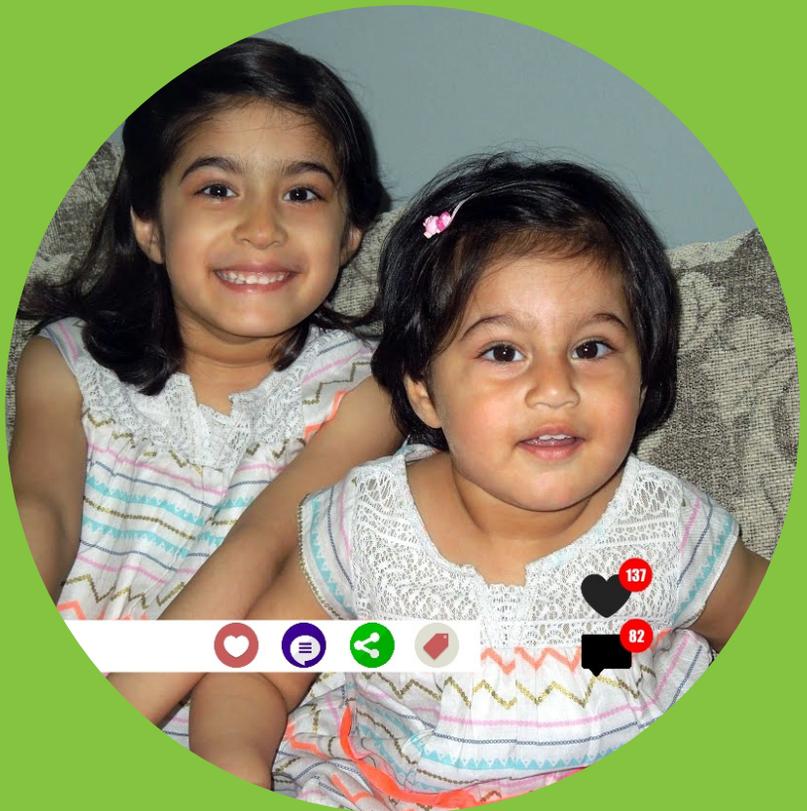

A! Aalto University

DOCTORAL DISSERTATIONS

# Underpinnings of Digital-photo interaction in Computer-mediated platforms

**Aqdas Amin Malik**

**Aalto University**
**School of Science**
**Deaprtment of Computer Science**
**Strategic Usability Research Group (STRATUS)**

**Supervising professor**
Professor Marko Nieminen, Aalto University, Finland

**Thesis advisor**
Dr. Amandeep Dhir, University of Helsinki

**Preliminary examiners**
Professor Eszter Hargittai, Department of Communication Studies, Northwestern University, U.S.A
Professor Heather Lipford, Department of Software and Information Systems, University of North Carolina at Charlotte, U.S.A

**Opponent**
Professor Anabel Quan-Haase, Faculty of Information and Media Studies, University of Western Ontario, Canada

NORDIC ECOLABEL

441    697
Printed matter

**Abstract**

Despite the growing popularity of digital photos in a number computer-mediated platforms, various pertinent issues remain less researched. Exploring and identifying relevant issues can help the researchers and designers understand "how and why SNS users interact with photos" on these platforms. In addition to this, the proposed investigations can be useful in refining the existing photos related features and solutions as well as developing new ones.

The present dissertation focuses largely on photo-sharing and photo-tagging activities, which constitute integral elements of digital-photo interaction in numerous computer-mediated platforms. This dissertation is based on five empirical research articles investigating the different latent factors that motivate and hinder the process of digital-photo interaction in computer-mediated platforms. Study I examine the current practices surrounding digital photos in the context of personal photo repositories (N=15). Study II investigates the gratifications and impeding factors associated with photo-tagging activity on Facebook (N=67). Study III develops and tests an instrument for understanding the gratifications of Facebook photo-sharing (N=368). Study IV examines the impact of various aspects of privacy in relation to photo-sharing intentions on Facebook (N=378). Finally, study V investigates the age and gender differences regarding various aspects of privacy and trust in the context of photo-sharing activity on Facebook (N=378).

The dissertation reveals the following findings: First, lack of "social features" is one of the essential reasons for non-acceptance of tagging feature in standalone photo management applications (Study I). Second, photo-sharing and photo-tagging adoption and popularity can be attributed to various factors such as affection, attention, communication, disclosure, habit, information sharing, self-expression, socialization, and social influence (Study II, III). Third, age, gender, and activity influence photo-sharing and photo-tagging gratifications (Study II, III). Fourth, in the context of Facebook photo-sharing, various aspects of privacy significantly impact users trust and activity and consequently photo-sharing intentions (Study IV). Fifth, women and young Facebook users are significantly more concerned about the privacy of their shared photos (Study V). Sixth, privacy-protective measures are significantly exercised more by young Facebook users, yet they exhibit more trust and a higher level of activity on Facebook (Study V). Overall, this dissertation work contributes to the existing theoretical and practical knowledge on activities associated with digital photos in computer-mediated platforms. Dissertation findings can be utilized by scholars engaged in researching computer-mediated communication, new media, privacy, and SNS behaviors.

**Keywords** Digital photos; Photo sharing; Photo tagging; Computer-mediated communication; Social Networking Sites; Facebook; Privacy

**Tiivistelmä**

Digitaalisten valokuvien käyttö viestinnän keinona on kasvanut viimeisten vuosien aikana erittäin paljon. Tässä väitöskirjassa tarkastellaan sitä, miksi ja miten sosiaalisen median palveluiden käyttäjät ovat vuorovaikutuksessa valokuva-aineiston kanssa ja kommunikoivat valokuvilla toistensa kanssa. Erityisesti tutkimuksen kohteena ovat valokuvien jakaminen ja linkittäminen, jotka muodostavat valokuvapohjaisen kommunikoinnin ytimen. Väitöskirja rakentuu viidessä julkaisussa raportoitujen tulosten ja analyysien varaan. Julkaisuissa tarkastellaan tekijöitä, jotka lisäävät ja vähentävät käyttäjien kiinnostusta ja innokkuutta valokuvien avulla kommunikoimiseen.

Työn ensimmäisessä julkaisussa (I) tarkastellaan digitaalisten valokuvien käsittelyyn liittyviä henkilökohtaisia käytäntöjä (N=15). Julkaisussa II kohdennutaan erityisesti tyytyväisyyttä ja tyytymättömyyttä aiheuttaviin tekijöihin Facebook-palvelussa tapahtuvaan valokuvien merkitsemiseen ja linkittämiseen liittyen (N=67). Väitöstyön kolmannessa julkaisussa (III) esitellään erityisen kyselyinstrumentin kehittäminen, jolla voidaan tutkia ja mitata sosiaalisessa mediassa tapahtuvaan valokuvien jakamiseen liittyviä mielihyvään vaikuttavia tekijöitä (N=368). Neljännessä julkaisussa (IV) keskitytään tarkastelemaan käyttäjien kokemien yksityisyyteen liittyvien tekijöiden vaikutusta valokuvien jakamisinnokkuuteen (N=378). Työn viimeisessä artikkelissa (V) analysoidaan lisäksi iän ja sukupuolen vaikutuksia yksityisyyden ja luottamuksen kokemiseen sosiaalisen median valokuvanjakotoimintoja käytettäessä (N=378).

Työn tuloksista selviää, että valokuvien merkitsemisen ja linkittämisen yksi keskeisimmistä syistä on valokuviin liittyvien ohjelmistojen ja palveluiden mahdollistama sosiaalinen vuorovaikutus (I). Valokuvien jakamiseen ja linkittämiseen liittyvien toimintojen käyttöön vaikuttavia tekijöitä ovat kiintymys, huomio, kommunikointi, julkistaminen, tottumus, tiedon jakaminen, itsensä ilmaiseminen, sosiaalistuminen ja sosiaalinen vaikuttaminen (II ja III). Ikä ja sukupuoli vaikuttavat tulosten mukaan valokuva-aineiston jakamisesta koettuun mielihyvään (II ja III). Facebook-palvelun käyttöön keskittyneessä tutkimuksessa havaittiin, että yksityisyyden kokemukseen liittyvät tekijät vaikuttavat merkittävästi käyttäjien innokkuuteen jakaa valokuva-aineistoa (IV). Naiset ja nuoremmat sosiaalisen median käyttäjät ovat viidennen tutkimuksen tulosten perusteella enemmän huolissaan valokuviensa yksityisyyteen liittyvistä seikoista (V). Nuoret käyttäjät näkevät enemmän vaivaa yksityisyytensä suojaamiseen, mutta kokevat samalla suurempaa luottamusta käyttämäänsä palveluun (V).

# Acknowledgements

My endeavor for doctoral research has been full of exciting and memorable times, which are characterized by novel learning, international exposure, collaboration, and humor. At this point, I am foremost grateful to the Almighty Allah and would also like to express gratitude to all the people who have supported me and whom I have had the honor to work with during the last four years of this remarkable journey.

**Honorable supervisor**
Professor Marko Nieminen

**The distinguished examiners**
Professor Eszter Hargittai
Professor Heather Lipford

**The distinguished opponent**
Professor Anabel Quan-Haase

**Research visit hosts**
Professor Donna Hoffman and Professor Thomas Novak
"Center for the Connected Consumer", George Washington University, USA

**Funders**
The Finnish Funding agency for Innovation (TEKES) - Data to Intelligence (D2I), Future Industrial Services (FutIS), & Service for fleet (S4Fleet) programs

Helsinki Doctoral Education Network in Information and Communication Technology (HICT)

**Collaborating organizations**
Nokia
F-Secure
PacketVideo
Tampere University of Technology

**Co-authors, colleagues and friends**
Amandeep Dhir
Kari Hiekkanen
Suleiman Ali Khan
Muhammad Ammad-ud-Din
Ali Faisal
Hussnain Ahmed
Iftikhar Ahmad
Asko Lehmuskallio
Ghufran Hashmi
Rao Anwer
Omer Anjum
Adnan Ghani
Mika Nieminen
Petri Mannonen
Sirpa Riihiaho
Junying Zhong
Sami Laine
Tapio Haanperä
Katrine Mahlamäki
& many others

**Family members**
Ami & Abu
Fahd & Umaira
Asdaq & Anza
Sana, Ali, & Ahmad
Bushra, Iman & Zainab

Otaniemi, Espoo, FINLAND, 5/5/16
Aqdas Amin Malik

This work is dedicated to my parents (Ami & Abu) for their endless love, support, and encouragement.

# Contents

# List of Original Publications

This doctoral dissertation consists of a summary and of the following publications which are referred to in the text by their numerals.

**I.** Malik, A., & Nieminen, M. (2014). Understanding the Usage and Requirements of the Photo Tagging System. Human IT 12.3: 117–161.

**II.** Malik, A., Dhir, A., & Nieminen, M. (2015). Facebook Photo Tagging Culture and Practices among Digital Natives. In CCGIDIS 2015-Fifth International Symposium on Communicability, Computer Graphics and Innovative Design for Interactive Systems, Madrid, Spain.

**III.** Malik, A., Dhir, A., & Nieminen, M. (2016). Uses and Gratifications of digital photo sharing on Facebook. Telematics and Informatics, 33(1), 129-138.

**IV.** Malik, A., Hiekkanen, K., Dhir, A., & Nieminen, M. (2016). Impact of Privacy, Trust, and User Activity on Intentions to Share Facebook Photos. Journal of Information, Communication & Ethics in Society, *(in press: vol: 14, iss: 3).*

**V.** Malik, A., Hiekkanen, K., & Nieminen, M. (2016). Privacy and trust in sharing photos on Facebook - Age and gender differences. (*submitted to a journal in March 2016).*

# Author's Contribution

**Publication I:** Understanding the Usage and Requirements of the Photo Tagging System.

As the first author, Aqdas Malik has developed the study plan, organized and implemented the user testing, and analyzed the study results. Aqdas has written the whole article with suggestions and comments from the co-author.

**Publication II:** Facebook Photo Tagging Culture and Practices among Digital Natives.

Aqdas Malik's contribution to this article is together with co-authors, the design, and analysis of the qualitative study. Aqdas has contributed significantly to writing the whole article with suggestions and comments from the co-authors.

**Publication III:** Uses and Gratifications of digital photo sharing on Facebook.

Aqdas Malik has been the main contributor, with support from the co-authors, to the design and implementation of the study. Aqdas has also supported co-authors in data analysis phase. Aqdas has written the whole article except the results section with suggestions and comments from the co-authors.

**Publication IV:** Impact of Privacy, Trust, and User Activity on Intentions to Share Facebook Photos.

Aqdas Malik has been the sole contributor to the design and implementation of the study. Aqdas has also supported co-authors during the data analysis phase. Aqdas has written the whole article except the results section with suggestions and comments from the co-authors.

**Publication V:** Privacy and trust in sharing photos on Facebook - Age and gender differences.

Aqdas Malik has been the sole contributor to the design and implementation of the study. Aqdas has also supported co-authors during the data analysis phase. Aqdas has written the whole article except the results section with suggestions and comments from the co-authors.

# 1. Introduction

The aim of this dissertation is to increase our understanding of digital-photo interaction i.e. the current practices associated with digital photos and how various factors influence them in computer-mediated platforms. The sharing and tagging of digital photos are two important and popular activities practiced on various computer-mediated platforms including Social Networking Sites (SNS). Over the past few years, there has been an exponential increase in the number of photos shared and tagged on various SNS platforms –notably Facebook. Similarly, personal and content privacy concerns associated with shared content (e.g. photos) have risen sharply among SNS users. Despite this interesting conflict, various SNS platforms are thriving with photos. Due to the increased popularity and adoption of digital photos as well as privacy issues associated with them, it is important to explore and understand the related phenomenon more deeply. By adopting a granular approach at a feature-specific level can be helpful in understanding some aspects associated with digital photos.

This dissertation aims to bridge the research gap by exploring and identifying some of the pertinent issues associated with digital photos on Computer-mediated platforms (specifically personal photo repositories and Facebook). The development of research models and empirical examination of the relationships between different attributes of digital-photo interaction were carried out to address the research questions of this dissertation. By employing Uses and Gratifications (U&G) theory and earlier literature on computer-mediated communication, online privacy, age and gender differences, and digital photography practices, this dissertation empirically examines the practices surrounding digital photos as well as the factors that impact these practices. To achieve the results, online surveys, open-ended questionnaire, and task analysis approaches were utilized to gather empirical data. Findings from the gathered empirical data formed the grounding of this dissertation work.

The next chapter introduces the key concepts, the theoretical framework utilized, and the importance of studying age, gender, and privacy in relation to digital photos. Subsequently, research gaps identified in the preceding litera-ture is presented. Finally, the chapter concludes by presenting the research framework and the main aims of this dissertation work.

## 1.1 What is Digital-Photo Interaction?

The popularity of Social Networking Services (SNS) has gained attention from various academic domains. An expanding body of research has explored different aspects of usage behaviors and motivations associated with these platforms. Digital photos are regarded as one of the most frequent and popular forms of content user interact on SNS platforms (Eftekhar, Fullwood, & Morris, 2014; Lee, Lee, Moon, & Sung, 2015; Pai & Arnott, 2013). A number of SNS features employ photos, and a multitude of activities are carried out around them, such as sharing, tagging, untagging, commenting, and liking photos (Dhir, Chen, & Chen, 2015; Lang & Barton, 2015; Tosun, 2012).

In the context of this dissertation, "digital-photo interaction" refers to the underlying usage and motivations of sharing and tagging photos. In addition to this, digital-photo interaction deals with the interplay of various elements (e.g. privacy, age, and gender) with the usage and motivations of sharing and tagging digital photos. For instance, on Facebook after uploading a photo, the user shares the photos with friends. During the sharing process, the user might tag a number of friends in the photo. Later on, the tagged photo shared on Facebook can get likes and a number of comments from other users. The same photo can also be untagged by the person who was tagged in the photo. Furthermore, the shared photo can be downloaded and re-shared by other users. This example illustrates some of the numerous interactions occurring around a single photo. These interactions are also coupled with various factors that can potentially influence photos-related activities.

Due to the increasing popularity and adoption of digital photos on various SNS platforms, scholars have recently initiated thorough investigations associated with these interactions (Dhir et al., 2015; Eftekhar et al., 2014; Litt & Hargittai, 2014; Wisniewski, Xu, Lipford, & Bello-Ogunu, 2015). Even though a wide array of features and activities encompass digital-photo interaction on SNS, this dissertation mainly focuses on interactions surrounding Facebook photo sharing and photo tagging activity and to some extent personal photo repositories (e.g. photos saved on personal computer/drives).

### 1.1.1 What is photo sharing?

Photo sharing is an act or process of transferring and publishing digital photos on a computer-mediated platform. This process enables the users to share his/her digital photos with the public or a selected group of people. E-mailing, media transfer (e.g. USB sharing), or online photo sharing platforms are some of the most common channels for sharing digital photos. Recent years have witnessed an ever-increasing popularity of photo-sharing activity on various computer-mediated platforms. For instance, almost 70% of all the Internet users in the U.K shared photos in 2013 (Dutton, Blank, & Groselj, 2013). Furthermore, engagement around digital photos has become the most popular leisure and entertainment activity on the Internet, superseding music listening, music downloads, games, and videos (Dutton et al., 2013). Due to possibilities of high interactivity and communicability with large (and limited) number of online audience, photo sharing feature is offered by almost all SNS platforms. Facebook -the most popular SNS utilized for maintaining and establishing new relationships is considered to be the most popular photo sharing SNS, as it has become the largest and fastest growing photo sharing platform (Rainie, Brenner, & Purcell, 2012). Despite the high adoption of photo sharing on Facebook, inquiries into the associated aspects have remained largely unexplored.

### 1.1.2   What is photo tagging?

Photo tagging is an act or process of labeling one's digital photos on a computer-mediated platform. The tagging feature has long been a primary feature of many standalone and online photo management and sharing platforms (Angus & Thelwall, 2010; Nov & Ye, 2010). However, the introduction of the tagging feature by Facebook has made this feature much more popular and acceptable (Pesce, Casas, Rauber, & Almeida, 2012). Even though Facebook tagging feature is highly limited as compared to traditional photo tagging features, it is one of the most novel, entertaining, and widely used features of the platform (Dhir et al., 2015). By tagging a photo on Facebook, a link to tagged person's profile is created. The link to the tagged person's profile is visible in form of his/her name when someone hovers over the tagged photo. Depending upon the settings, the tagged photo may also appear on the tagged person's timeline. All the tagged users in a photo are subsequently notified via an alert message. Tagging one's photo on Facebook not only increases the visibility but also the possibility of being re-shared (Besmer, Lipford, & Lipford, 2010). Until now, little research has examined attitudes and behavioral aspects associated with this feature, though some research has been carried out on understanding privacy implications associated with Facebook tagging (Besmer et al., 2010; Pesce et al., 2012; Wisniewski et al., 2015). More recently, in the context of managing undesirable Facebook photos, untagging behavior has also been studied (Dhir, Kaur, Lonka, & Nieminen, 2016; Lang & Barton, 2015).

## 1.2   What is Uses and Gratifications (U&G) theory?

Uses and Gratifications theory (U&G) aims to explain the uses and functions of media for individuals, groups, and society at large (Katz, Haas, & Gurevitch, 1973). The U&G framework attempts to fulfill three main objectives: a) explaining how people use media for gratifying their needs, b) discovering the fundamental motivations for the use of media, and c) identifying the positive and negative consequences arising from the media usage (Katz et al., 1973). U&G assumes that the individuals are goal-oriented and are active in seeking and using the media that best gratify their specific needs. Furthermore, if those needs are fulfilled; the people are more likely to use and continue the usage of that media (Katz et al., 1973).

  Relatively recent research has widely adopted U&G framework to understand the motives for the usage of SNS. All the SNS target different demographics and offer a wide array of features, resulting in distinctive usage behavior, motivations, and interactions (Quan-Haase & Young, 2010; Sundar & Limperos, 2013). Moreover, practices and motivations associated with usage of different features across SNS are also not uniform (Karnik, Oakley, Venkatanathan, Spiliotopoulos, & Nisi, 2013; Smock, Ellison, Lampe, & Wohn, 2011). Due to these variations, gratifications and usage of a particular SNS might not be generalizable and applicable to other SNS as such. Likewise, feature-specific gratifications also vary substantially among each SNS, hence limiting the generalizability of feature-specific characteristics of a particular SNS. Based on the above premise, it is necessary to approach the behavior and practices of SNS users by unbundling specific features of each SNS (Baek, Holton, Harp, & Yaschur, 2011; Karnik et al., 2013; Smock et al., 2011). By studying feature-specific usage behavior and relevant aspects, highly interesting trends, user perceptions, and novel usage patterns relevant to specific features can be identified (Karnik et al., 2013; Smock et al., 2011).

## 1.3   Important of understanding age and gender differences in SNS

Age and gender variables are regarded as one of the central determinants that influence the usage and adoption of the various computer-mediated platforms, including collaborative virtual environments (Felnhofer et al., 2014), online shopping (Lian & Yen, 2014), and general Internet use (van Deursen & Van

Dijk, 2014). SNS attract users from diverse demographic backgrounds and their usage, motivations, and related behavior differs substantially among different groups. Demographic understanding and investigations especially around age and gender are two of the highly important correlates of SNS usage (Kimbrough, Guadagno, Muscanell, & Dill, 2013). Understanding these differences can help designers and developers to create and calibrate applications and associated features accordingly. Similarly, companies promoting their products/services to a specific group of SNS users can make well-informed decisions. Prior studies highlight interesting relationships and impacts of age and gender on specific activities, frequency, and usage of SNS. For instance, young SNS users mostly engage with these platforms for habitual purposes and as a pastime activity (Papacharissi & Mendelson, 2010), meanwhile, older SNS users use it for connecting with their families, acquaintances, and colleagues (Bell et al., 2013; Brandtzæg, Lüders, & Skjetne, 2010). Older users have a limited network of friends while young SNS users have a much wider friends network (Bell et al., 2013; Brandtzæg et al., 2010; Hope, Schwaba, & Piper, 2014). Due to different motivations and communication needs, behaviors and usage patterns of different age groups vary substantially, therefore, it is important to understand these differences from the perspective of digital-photo interaction.

   With respect to gender, significant differences also exist between males and females as far as SNS usage and behavior are concerned. For instance, majority of the female SNS users engage with the platform as a pastime, to entertain themselves, and to communicate with others (Hunt, Atkin, & Krishnan, 2012; Sheldon, 2008), meanwhile most male SNS users seek new relationships and romance on these platforms (Sheldon, 2008). Compared to men, women are more attached and highly active in various social activities and tend to have a higher number of friends on SNS (Kimbrough et al., 2013; McAndrew & Jeong, 2012). Moreover, female SNS users spend far more time browsing through other's profiles and photos, while male users are more active in playing social games (Muscanell & Guadagno, 2012). Based on these above stated differences, we assume that digital-photo interaction also differs among these groups and require further investigation.

## 1.4   Important of studying SNS privacy aspects

 The pervasive nature of many computer-mediated platforms often leads to unintended consequences such as privacy threats. Various aspects of online privacy have been studied in the context of a number of online platforms in-cluding online shopping (Nepomuceno, Laroche, & Richard, 2014), blogging (Viégas, 2005), mobile commerce (Zhang, Chen, & Lee, 2013), and email (Hornung, 2005). In recent years, privacy has focused around SNS, as these platforms become highly popular among online users. The majority of the SNS are developed with the aim to encourage users to disclose various forms of content including photos, videos, location, and personal information. Providing information on SNS can have various privacy consequences such as cyber bullying, damaged reputation, identity theft, and harassment (Christofides, Muise, & Desmarais, 2012; Debatin, Lovejoy, Horn, & Hughes, 2009; Lipford, Besmer, & Watson, 2008). Enormous acceptance and adoption of SNS call for increased understanding of privacy attitudes and behaviors of the users. As the recent literature indicate, SNS privacy issues are highly complicated to comprehend, hence understanding them through a nuanced approach might lead to simplified insights.

## 1.5   The present study

In the proceeding sub-section, research gaps and limitations identified in the previous literature coupled with briefings about the contributions of this dissertation are presented. Later, the research framework utilized in this work is described. Finally, the main aims of this dissertation work are outlined in the concluding sub-section.

### 1.5.1   Research Gaps and contributions of this dissertation

Enormous adoption of SNS has led to extensive research on understanding the behaviors and motivations of SNS users, yet some limitations and research gaps still prevail in the current literature. Considering this impetus and research paucities, this dissertation aims to contribute empirically by complementing and addressing these limitations and gaps.

First and foremost, activities surrounding digital photos are highly popular on various computer-mediated platforms. Despite being one of the frequently interacted and widespread content on a number of SNS, limited research has addressed issues pertaining to digital photos on SNS platforms. A number of photo-related activities and features are highly interconnected, and various factors play a role in promoting or demoting them. Hence, a refined understanding of this linkage and the influence of various factors are important to grasp the context of photo related activities on SNS. To address this research gap, the current dissertation work attempts to bring a better understanding of various features and factors associated with digital photos on SNS. To com-prehend and provide clarity to this phenomenon, the present study has introduced the term "digital-photo interaction."

Second, there is an expanding body of research examining various aspects of usage behaviors on different SNS platforms (boyd & Hargittai, 2010; Papacharissi & Mendelson, 2010; Quan-Haase & Young, 2010). Despite these insightful examinations hailing from various domains, limited research has explored usage behaviors of specific features offered by these platforms (Krause, North, & Heritage, 2014; Smock et al., 2011). To address this limitation, the present study specifically concentrated on two highly popular elements of digital-photo interaction i.e. photo sharing and photo tagging (Study I-V).

Third, despite being one of the leading activities on various SNS, only limited research is available which empirically examines factors that motivate users to engage with photo-sharing activity on these platforms (Bayer, Ellison, Schoenebeck, & Falk, 2015; Sheldon & Bryant, 2016; Thelwall et al., 2015). This dissertation work fills this gap by developing and testing a U&G instrument to identify various gratifications of photo-sharing activity on Facebook (Study III).

Fourth, recent literature that examines U&G of particular Facebook features has predominantly focused on adults, particularly college/university students (Baek et al., 2011; Karnik et al., 2013). Despite being one of the predominant group using SNS – the adolescent users have been rarely studied (Kaur, Dhir, Chen, & Rajala, 2016a, 2016b). Moreover, a limited amount of research has been carried out with SNS users in developing nations as the majority of the research on U&G of SNS has been carried out in a Western context (Chang & Heo, 2014; Quan-Haase & Young, 2010). This dissertation work addresses these gaps by utilizing a qualitative study to examine the practices, motivations, and impediments of engaging in Facebook photo tagging activity by adolescent Facebook users in India (STUDY II).

Fifth, prior literature on computer-mediated platforms considers age and gender as highly important determinants of technology use and adoption (van Deursen & Van Dijk, 2014; Weiser, 2000). Recent studies on SNS behaviors also highlight the importance of these factors in determining usage, motivations, and communication patterns (Hargittai, 2007; McAndrew & Jeong, 2012; Muscanell & Guadagno, 2012). Despite being highly important factors influencing SNS usage and adoption, age and gender have not been studied thoroughly at a feature specific level (e.g. photo-related features). To address this gap, this dissertation work addresses the influence of age and gender on various aspects that relate to sharing photos on Facebook (Study V). In addition, age and gender differences with respect to photo sharing gratifications were also examined (Study III).

Sixth, online privacy has been a highly active research topic as it impacts user engagement and adoption in computer-mediated platforms (Nepomuceno et al., 2014; Zhang et al., 2013). Various aspects of privacy have also received deserved attention from scholars exploring SNS behaviors (boyd & Hargittai, 2010; Debatin et al., 2009; Wisniewski et al., 2015; Young & Quan-Haase, 2013). However, due to the comparatively new communication medium and rapidly evolving nature of SNS, there is a need to extend and refresh current understanding on various aspects of SNS privacy, especially from the perspective of particular features or content type. Digital photos offer high potential for misuse and abuse that can lead to negative consequences e.g. cyber bullying, damaged reputation, and identity theft (Christofides et al., 2012; Debatin et al., 2009; Lipford et al., 2008). Hence, it is critical to understand SNS privacy and its relationship with digital photos further. The current study addresses this gap by developing and empirically validating a model to uncover the relationship of various privacy aspects with trust, Facebook activity, and photo-sharing on Facebook (Study IV). Furthermore, the current study also explored age and gender differences on various aspects of privacy associated with photo-sharing on Facebook (Study V).

### 1.5.2   A summary of the Research Framework

The research framework (see Figure 1) exemplifies the research questions and main aims of this dissertation. The key impetus behind this framework is to conceptualize clearly and highlight some of the important attributes of digital-photo interaction in computer-mediated platforms. As described earlier, digital-photo interaction encompasses a wide array of features and activities (e.g. photo sharing and photo tagging). Furthermore, these features and surrounding activities are influenced by numerous factors such as privacy, trust, and demographic characteristics. Recent literature on SNS behaviors has emphasized the significance of adopting a micro approach to gain a better understanding of feature usage and associated factors. Therefore, examining "digital-photo interaction"- various features and activities related to photos as well as the factors that impact their usage and adoption is necessary. The research framework proposed in this study particularly concentrates on the two most popular features of digital-photo interaction i.e. photo sharing and photo tagging. Even though the present study does not accommodate other features and activities presented in the research framework, they are integral and complimenting elements of digital-photo interaction. Likewise, the current research framework accommodates limited factors that influence digital-photo interaction. In practice, there might be a multitude of other factors that are not addressed by this dissertation (e.g. cultural influence, educational background, and tenure on SNS).

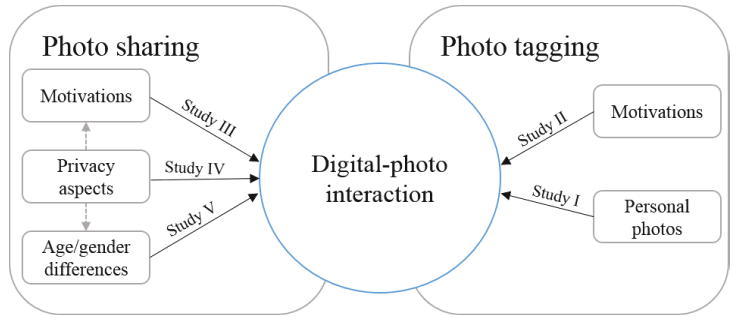

**Figure 1.** Overview of the research framework

### 1.5.3 Main aims

The current dissertation work embodies a multi-disciplinary perspective on the conceptualization and nuanced understanding of digital-photo interaction in computer-mediated platforms. In pursuit of addressing the underlying research questions, concepts and literature from several domains, including computer-mediated communication, human-computer interaction, new media, and psychology, were utilized. Due to the rapidly changing nature of SNS and the complexities associated with the behaviors of SNS users, understanding and quantifying these behaviors in general is not adequate. Instead, the behaviors of SNS users can be better understood by adopting granular approaches. Examining the influences and impacts of numerous factors such as demographic attributes (i.e. age and gender), privacy impediments, trust, and sought U&G associated with a feature or activity can lead to novel and insightful feature/activity specific perspectives. Consequently, this dissertation aims at providing an in-depth illustration of some aspects of digital-photo interaction on computer-mediated platforms.

Accordingly, this dissertation achieves these aims by addressing the following research questions:

**RQ1:** What are the different practices associated with the personal photo repositories (Study I)?

**RQ2:** Why do people tag photos on SNS (Study II)?

**RQ3:** Why do people share photos on SNS (Study III)?

**RQ4:** How various aspects of privacy influence photo-sharing activity on SNS (Study IV)?

**RQ5:** How does age and gender differ with respect to various privacy aspects associated with SNS photo-sharing (Study V)?

Each of the five research articles, which are independent studies, do not specifically address all aspects of the current study. Instead, these empirical studies complement each other to fulfill the aims of this dissertation. Additionally, this work cannot be considered comprehensive as it addresses only some aspects of digital-photo interaction.

**Table 1.** Summary of five empirical studies

| Studies | Main aims | Participants | Measures | Analysis |
|---|---|---|---|---|
| Study I | To examine the current practices and user needs of photo tagging practices in personal photo repositories | 15 test participants (aged 18-39 years) | Imaging practices and 9 test tasks | Affinity diagramming |
| Study II | To examine the gratifications and practices associated with photo tagging activity on Facebook | 67 adolescent Facebook users aged 13-18 years | 8 open-ended questionnaire | Affinity diagraming |
| Study III | To investigate the uses and gratifications associated with photo-sharing activity on Facebook | 368 Facebook users (aged over 18 years) | 26-item photo sharing U&G instrument | Exploratory factor analysis, Correlation analysis, Independent samples t-test |
| Study IV | To understand the impact of various privacy aspects, trust and user activity on users' intentions to share photos on Facebook | 378 Facebook users (aged over 18 years) | Privacy concerns, privacy-seeking behavior, privacy awareness, trust, and sharing intentions | Partial least squares path modeling (PLS) |
| Study V | To examine the impact of age and gender on various aspects of privacy and trust in context of photo sharing on Facebook | 378 Facebook users (aged over 18 years) | Privacy concerns, privacy-seeking behavior, privacy awareness, trust, and sharing intentions | Independent-samples t-test and Analysis of variance (ANOVA) |

# 2. BACKGROUND LITERATURE

This chapter reviews significant relevant work regarding uses and gratifications (U&G) of Social networking sites (SNS) and their specific features, digital-photo interaction in various computer-mediated platforms, particularly SNS (photo sharing and photo tagging), and the factors that impact digital-photo interaction such as age, gender, and privacy.

## 2.1 Uses and Gratifications (U&G) of SNS

Recently, new media especially web-based technologies and services have been increasingly studied by utilizing the U&G approach. Due to the highly interactive nature of web-based technologies/services and particularly computer-mediated platforms, U&G is considered highly appropriate and valuable for studying them (Ruggiero, 2000). Prior extended literature examining the U&G of computer-mediated platforms revealed different sets of motivations behind their use. These included information seeking, entertainment, time passing, interpersonal communication, convenience (Papacharissi & Rubin, 2000), relationship maintenance, status seeking, personal insights, problem-solving, persuading others (Flanagin & Metzger, 2001), interactivity, economic motivations (Korgaonkar & Wolin, 1999), virtual community gratification (Song, Larose, Eastin, & Lin, 2004), and search and cognitive factors (Stafford & Stafford, 2002).

More recently, the popularity of different SNS has led to the detailed examination of different U&G offered by these platforms. These investigations mainly focused on Facebook and Myspace (Papacharissi & Mendelson, 2010; Quan-Haase & Young, 2010; Smock et al., 2011), Pinterest (Mull & Lee, 2014), Twitter (Chen, 2011; Johnson & Yang, 2009), and LinkedIn (Florenthal, 2015). These studies have made significant theoretical contributions to new media and U&G literature by empirically identifying a number of new gratifications that were non-existent earlier. For instance, Mull and Lee (2014) examined the U&G of Pinterest and found five U&G namely creative projects, virtual exploration, entertainment, fashion, and organization. Some of these U&G were not identified in prior SNS literature. Similarly, U&G of SNS use including expression of opinion, coolness, knowledge/surveillance about others, and convenience are also novel motivations (Hollenbaugh & Ferris, 2014; Sheldon & Bryant, 2016; Whiting & Williams, 2013). To understand the U&G offered by different SNS, a thorough review of the prior literature was performed (see Table 2).

**Table 2.** U&G of different SNS

| Author(s) | SNS Platform (context) | Methodology | Gratifications |
|---|---|---|---|
| (Pai & Arnott, 2013) | Facebook | 24 Interview with Taiwanese Facebook users | Belonging, hedonism, self-esteem, reciprocity |
| (Al-Jabri, Sohail, & Ndubisi, 2015) | Twitter | Online survey with 281 Saudi Twitter users | Enjoyment, freedom of expression, social interaction |
| (Pentina, Basmanova, & Zhang, 2016) | Twitter | Online survey with 125 Ukrainian and 77 US-based Twitter users | Professional development, entertainment, status maintenance, social interaction, self-expression, news updates/sharing |
| (Whiting & Williams, 2013) | Social Media (In general) | 25 interviews with SNS users | Social interaction, information seeking, pass time, entertainment, relaxation, communication, expression of opinion, information sharing, knowledge about others |
| (Cheng, Liang, & Leung, 2014) | SNS (usage on mobile devices) | Survey with 760 university students in China | Accessibility, affection, cognition needs, recognition, fashion/status |
| (Mo & Leung, 2015) | Sina Weibo | Online survey with 431 Weibo users in China | Social (to meet new friends, to get to know people with shared interests), content (to solicit information, to share information with others) |
| (Sheldon & Bryant, 2016) | Instagram | Survey with 239 students at an American university | Surveillance/knowledge about others, documentation, coolness, creativity |
| (Hicks et al., 2012) | Yelp | Online survey with 144 Yelp users in USA | Information-seeking, entertainment, convenience, interpersonal utility, pass time |
| (Lee et al., 2015) | Instagram | Online survey with 212 Korean Instagram users | Social interaction, archiving, self-expression, escapism, peeking |
| (Quan-Haase & Young, 2010) | Facebook | Survey with 77 participants and 21 interviews with Canadian University students | Pastime, affection, fashion, share problems, sociability, social information |

| (Doty & Dworkin, 2014) | SNS (for parenting purpose) | Online survey with 649 parents of adolescents in USA | Communicate with their child/extended family members/children's friends/ parents of their children's friends, connect with community |
|---|---|---|---|
| (Alhabash, Park, Kononova, Chiang, & Wise, 2012) | Facebook | Online survey with 4,346 Taiwanese Internet users | Social connection, shared identities, content, photographs, social investigation, status updates |
| (Papacharissi & Mendelson, 2010) | Facebook | Online survey with 334 US-based University students | Expressive information sharing, habitual pass time, relaxing entertainment, cool and new trend, companionship, professional advancement, escape, social interaction, new friendships |
| (Park & Lee, 2014) | Facebook | Online survey with 292 US-based University students | Entertainment, relationship maintenance, self-expression, communication, impression management |
| (Giannakos, Chorianopoulos, Giotopoulos, & Vlamos, 2013) | Facebook | Online survey with 222 Facebook users at a Greek University | Wasting time, social connection, social surfing, using applications |
| (Smock et al., 2011) | Facebook | Online survey with 267 US-based University students | Relaxing entertainment, expressive information sharing, escapism, cool and new trend, companionship, professional advancement, social interaction, habitual pass time, to meet new people |
| (Hollenbaugh & Ferris, 2014) | Facebook | Online survey with 301 random Facebook users | Relationship maintenance, passing time, virtual community, entertainment, coolness, companionship |

The review of prior literature examining U&G of SNS use revealed different set of U&G namely affection (Cheng et al., 2014; Quan-Haase & Young, 2010), entertainment (Hollenbaugh & Ferris, 2014; Park & Lee, 2014; Smock et al., 2011), pastime (Hollenbaugh & Ferris, 2014; Papacharissi & Mendelson, 2010; Quan-Haase & Young, 2010), information sharing and seeking (Papacharissi & Mendelson, 2010; Smock et al., 2011; Whiting & Williams, 2013), relationship maintenance (Hollenbaugh & Ferris, 2014; Park & Lee, 2014), communication (Alhabash et al., 2012; Doty & Dworkin, 2014; Park & Lee, 2014), self-expression (Lee et al., 2015; Whiting & Williams, 2013), social interaction (Giannakos et al., 2013; Quan-Haase & Young, 2010), and surveillance/knowledge (Sheldon & Bryant, 2016; Whiting & Williams, 2013).

### 2.1.1 Uses and Gratifications (U&G) of SNS features

In recent years, scholars have argued the need to further investigate the U&G of specific features of SNS (Dhir et al., 2015; Karnik et al., 2013; Smock et al., 2011). According to Dhir et al. (2015), the U&G of a specific SNS feature vary substantially from the U&G of other SNS features. Considering the importance of nuanced understanding with respect to specific features of SNS, fewer prior empirical studies (see Table 3) have examined the U&G of news sharing (Lee & Ma, 2012), music listening (Krause et al., 2014), brand pages (Tsai & Men, 2013), sharing links (Baek et al., 2011), and participation in groups (Karnik et al., 2013; Park, Kee, & Valenzuela, 2009) on SNS.

As briefed in the previous chapter, features and activities surrounding photos are becoming increasingly popular among SNS users, study II and study III were carried out to understand U&G of photo tagging and photo sharing activities on Facebook. These studies complement to the growing literature presented above that focuses on exploring U&G of specific features on SNS.

**Table 3.** Feature-specific U&G of SNS

| Author(s) | Platform and features | Methodology | Gratifications |
|---|---|---|---|
| (Lee & Ma, 2012) | News sharing in social media | Survey with 203 University students in Singapore | Information seeking, socializing, status seeking |
| (Krause et al., 2014) | Music listening applications on Facebook | Online survey with 153 participants | Entertainment, communication, habitual diversion |
| (won Kim, 2014) | Social recommendation systems | Online survey with 541 American University students | Expression, information, socialization, entertainment |
| (Tsai & Men, 2013) | Facebook brand pages | Online survey 280 with Facebook users in the United States | Remuneration, seeking product/brand/company-related information, entertainment |
| (Cheng & Leung, 2015) | Mobile social game | Online survey with 409 Candy Crush Saga users in China | Mobility, entertainment, sociability, achievement, relaxation |
| (Park et al., 2009) | Facebook groups | Online survey with 1,715 students in two US universities | Socializing, entertainment, self-status seeking, information |
| (Karnik et al., 2013) | Music video sharing groups on Facebook | Online survey with 57 participants | Contribution, discovery, social interaction, entertainment |

| (Baek et al., 2011) | Sharing links through Facebook | Online survey with 217 Facebook users | Information sharing, convenience, entertainment, pass time, interpersonal utility, control, promoting work |
|---|---|---|---|
| (Malik, Dhir, & Nieminen, 2016) | Photo sharing on Facebook | Online survey with 368 Facebook users | Affection, attention seeking, disclosure, habit, information sharing, social influence |
| (Dhir et al., 2015) | Photo tagging on Facebook | Survey with 499 Indian Facebook users | Likes and comments, social influence, peer pressure, gains popularity, entertainment, feels good, social sharing, affection, convenience |

## 2.2   Digital-photo interaction

During the last decade, photographic practices such as capturing, storing, and disseminating photos have witnessed a comprehensive transformation due to wider adoption of digital cameras. Decreasing associated costs, quick results, no necessity of printing, and high storage capabilities fueled the adoption of digital photography in households (Kirk, Sellen, Rother, & Wood, 2006; Rodden & Wood, 2003). Digital photos also gave the freedom to take multiple photos of the same scene in a hope that one will be excellent (Kirk et al., 2006). Kirk et al. (2006) coined the term "photowork" encompassing various activities e.g. reviewing, downloading, organizing, editing, sorting and filing digital photos. Their study produced a workflow by observing the interactions of amateur photographers with digital photos. This workflow outlined a detailed description of motivations and management strategies associated with each workflow activity.

In recent years, digital-photo interaction has increased exponentially. Digital photos have become one of the most popular form of content interacted in a number of computer-mediated platforms. The sharing of digital photos online helps users to disseminate information to larger web audiences with ease (Naaman, Nair, & Kaplun, 2008). Furthermore, sharing photos online is regarded as one of the key drivers for photography (Ames, Eckles, Naaman, Spasojevic, & House, 2010). The prior empirical work carried out to understand the usage and motivations of digital photos have been mostly qualitative in nature. The majority of these studies were conducted with a small number of participants and predominantly studied Flickr.

The motivations of sharing photos online have been mainly categorized into intrinsic and extrinsic motivations. The intrinsic motivations for sharing pho-tos on Flickr are primarily enjoyment and commitment to the online community while extrinsic motivations include self-development and reputation (Nov, Naaman, & Ye, 2009, 2010). Motivations such as self-presentation, communication, and establishing and maintaining social relationship are some of the other prime extrinsic motivations for sharing photos online (Goh, Ang, Chua, & Lee, 2009; Oeldorf-Hirsch & Sundar, 2010; Van House, Davis, Ames, Finn, & Viswanathan, 2005). Moreover, habit, self-expression, entertainment, and self-disclosure are some of the other intrinsic motivations of sharing photos online (Van House, 2007; Miller & Edwards, 2007; Stefanone & Lackaff, 2009).

Prior literature also suggest that the motivations of sharing photos online are driven by functional purposes e.g. assisting personal and group tasks activities (Goh et al., 2009; Van House, 2007). In addition to this, digital photos are shared online with an intention to gain attention, feedback, recognition and social rewards from friends, family members, and other online users (Malinen, 2011; Oeldorf-Hirsch & Sundar, 2010). A list of online photo sharing motivations identified by prior literature is presented in Table 4.

**Table 4.** Motivations of sharing digital photos online

| Author(s) | Media | Method and sample | Motivations |
|---|---|---|---|
| (Van House et al., 2005) | Cameraphone | Interviews, focus groups, photo and examinations with 60 participants | Personal/group memory, creating/ maintaining social relationships, self-expression, self-presentation, functional purpose |
| (Van House 2007) | Flickr | Interviews and photo elicitation with 12 Flickr users | Memory, identity, narrative, maintaining relationships, self-representation, self-expression |
| (Goh et al., 2009) | Cameraphone | Diary study and interview with 18 students | Creating and maintaining social relationships, emotional/social influences, reminders, self-presentation, task performance, self-expression |
| (Nov et al., 2009) | Flickr | Survey with 278 respondents and independent system data | Enjoyment, commitment, self-development, reputation, structural embeddedness |
| (Oeldorf-Hirsch & Sundar, 2010) | Online photo sharing | Focus groups with 22 students. Online survey with 460 students | Seeking and showcasing experiences, website affordances/technological reasons, social connection/bonding, reaching out/bridging |

In addition to sharing photos, empirical work has also been carried out on photo tagging, which is also an important elements of digital-photo interaction in several computer-mediated platforms. Tagging is defined as the process of associating a label, term, annotation, or a "tag" with digital content, e.g. digital photo, video, document, blog, or bookmark (Nov & Ye, 2010). Tags are usually keywords about the content that provide additional information about the content (Ames & Naaman, 2007; Angus & Thelwall, 2010). A number of computer-mediated platforms, including notably Flickr, offer photo tagging feature with the aim of effectively organizing, searching, and sharing photos with other users (Ames & Naaman, 2007). Tagging photos on Flickr, which is usually termed as "Social tagging," helps the content owner as well as other users in fulfilling information management, self-organization, rediscovery, and attention-seeking needs (Angus & Thelwall, 2010).

Previous research on photo tagging has predominantly concentrated on developing and testing new tagging concepts and algorithms (Naaman & Nair, 2008). Similarly, the motivations and user needs of photo tagging have mostly been addressed in the context of the online photo management and sharing platform Flickr that offers limited SNS features (Ames & Naaman, 2007; Angus & Thelwall, 2010; Nov et al., 2009).

### 2.2.1 Digital-photo interaction on SNS

Digital photos are one of the most frequently interacted content by SNS users. Various activities related to digital photos such as viewing, uploading, sharing, tagging and un-tagging photos have become increasingly popular among SNS users (Eftekhar et al., 2014; Malik, Dhir, & Nieminen, 2015; Orito, Fukuta, & Murata, 2014; Wisniewski et al., 2015). Most SNS enable and motivate their users to share profile pictures, personal photos, and re-share photos posted by other users. In recent years, sharing photos especially on SNS has risen substantially, making it one of the most popular activities on these platforms (Duggan, 2013; Joinson, 2008; Pai & Arnott, 2013). More than half of internet users in the USA have shared photos online they captured themselves, while around 42% of them have forwarded or shared photos posted by others (Duggan, 2013). Adoption of digital photos can be witnessed by the fact that 90% of SNS teenage users share or forward photos through their SNS accounts (Madden, Lenhart, Duggan, Cortesi, & Gasser, 2013). Similarly, a UK-based study reported that engaging with photos is the leading activity on SNS, and more than 64% of British SNS users are involved in photos related activities (Dutton et al., 2013).

Due to ever-increasing popularity and usage of digital photos, research has recently begun to understand the usage, behaviors, and concerns related to these activities on various SNS platforms including Instagram, Facebook, and Snapchat (Bayer et al., 2015; Eftekhar et al., 2014; Lee et al., 2015). Most SNS users' predominantly share photos that depict positive things about themselves with an aim of managing their self-impressions (Dorethy, Fiebert, & Warren, 2013). Photos are also shared with the intention to start a conversation or an interactive discussion around them (Dorethy et al., 2013). Appraising self-worth, interest in photography, and showcasing their work are some of the other motivations of SNS-based photo sharing (Stefanone, Lackaff, & Rosen, 2010). SNS users select SNS profile photos to project the desired self-image, showcase friendship and relationship, and to represent a special occasion (K. Young, 2009). As various SNS host varying features, the motivations and usage of these features vary across the platforms. For instance, enjoyment, sharing selfies and spontaneous experiences, flirting, and procrastination are some of the main motivations of sharing photos on Snapchat (Bayer et al., 2015; Utz, Muscanell, & Khalid, 2015). On Twitter, photos are depicted as the visual extensions of chatting activity and the main motivation of sharing them is to update others and provide wider information sharing (Thelwall et al., 2015). Meanwhile, surveillance, documentation, status-seeking, and self-representations are some of the main gratifications of sharing photos on Instagram - a highly popular photo sharing SNS (Lee et al., 2015; Sheldon & Bryant, 2016). A detailed list of motivations for sharing photos on various SNS platforms identified by recent literature is presented in Table 5.

**Table 5.** Motivations of photo sharing on SNS

| Author(s) | Media | Method and sample | Motivations |
|---|---|---|---|
| (Bayer et al., 2015) | Snapchat | Online survey with 154 US-based University students. Interviews with 28 Snapchat users | Enjoyment, sharing mundane experiences with close ties and reduced self-presentational concerns, attention, sharing spontaneous experiences with trusted ties, sharing selfies |
| (Utz et al., 2015) | Snapchat | Online survey with 77 Snapchat users | Flirting, finding new love, communicating with partner, keeping in touch with, friends and family, seeing what people are up to, procrastination and distraction |

| (Thelwall et al., 2015) | Twitter | Content analysis of 800 random photos in UK and USA | Selfies, update friends, visual extension of chatting face-to-face, broadcast non-photographic visual status updates, wider information sharing |
|---|---|---|---|
| (Sheldon & Bryant, 2016) | Instagram | Survey with 239 students at an American university | Surveillance/knowledge about others, documentation, coolness, creativity |
| (Lee et al., 2015) | Instagram | Online survey with 212 Korean Instagram users | Social interaction, archiving, self-expression, escapism, peeking |
| (Hunt, Lin, & Atkin, 2014) | Photo-messaging | Online survey with 682 US-based university students | Self-presentation, self-expression, relationship maintenance, relationship formation |

Facebook has been recognized as the largest and the fastest growing photo-sharing SNS with over 1.44 billion monthly active users ("Company Info | Facebook Newsroom," 2015). By 2013, Facebook users had already shared 250 billion photos on the platform and every day they contributed 350 million photos (Internet.org, 2013). Likewise, on Instagram (recently acquired by Facebook), that has more than 200 million active users, 20 billion photos have been shared, and on average 60 million photos are shared every day ("Press Page - Instagram," 2016). Facebook users engage with digital photos for a number of reasons, primarily self-disclosure and self-presentation (Dhir 2016; Li-Barber, 2012; Ong et al., 2011). Many of the previous investigations on self-disclosure and self-presentation behaviors and practices have predominantly addressed behaviors and practices of textual content (status updates, personal information, and wall postings, etc.) on Facebook (Hollenbaugh & Ferris, 2014; Li-Barber, 2012). Limited research has been carried out exploring the behavioral aspects of visual content which is a popular form of self-disclosure and self-presentation on the platform. In recent years, there has been a growing interest among the research community to address this gap, as most of the earlier research on digital photos on Facebook has been briefly addressed as a subset of larger studies. Study II – V specifically address this research gap by examining various aspects associated with photo sharing and tagging activities on Facebook.

### 2.2.2 Photo sharing on Facebook

Sharing photos is one of the most popular forms of content and a leading activity on Facebook (Duggan, 2013; Madden et al., 2013). Facebook allows its users to share their photos with a huge network of users as well as with restric-tive access to their friends and family members (Rainie et al., 2012). The phenomenon of instant communication and feedback in the form of likes and comments is also considered highly gratifying and attributed to the popularity of Facebook photo sharing feature (Kumar & Schoenebeck, 2015; Madden et al., 2013). Sharing photos on Facebook has various motivational aspects. For instance, it helps the users in highlighting social relationships as well as in fostering social connections (Mendelson & Papacharissi, 2010). Photos are also shared on Facebook to thank someone publicly for a gift by tagging the photo to gift giver's profile (Kumar & Schoenebeck, 2015). Visual communication around shared photos on Facebook is also considered a crucial element for self-presentation and impressions management (Eftekhar et al., 2014; Siibak, 2009).

Research has also been carried out to gain insight into various predictors of photo-sharing activity on Facebook including age, gender, personality, and cultural differences. Facebook profile photos significantly impact the willingness of other users to initiate a friendship. Profile photos that portray physical attractiveness are more likely to receive friendship requests from other Facebook users (Wang, Moon, Kwon, Evans, & Stefanone, 2010). Narcissistic Facebook users choose physically appealing, attractive, and fashionable profile photos for the sake of self-presentation and self-promotion (Ong et al., 2011). Similarly, a study on Australian Facebook users suggests that users scoring high on exhibitionism prefer engaging with photo related activities on Facebook more often (Ryan & Xenos, 2011). A relatively recent study revealed that Facebook users displaying high openness and extroverted personalities tend to share more photos on the platform (Kuo & Tang, 2014). Another recent study by Eftekhar et al. (2014) revealed the impacts of the big five personality traits on photo sharing activity. The results indicate that extroverted and neurotic Facebook users share more photos as well as profile pictures. Meanwhile, agreeable Facebook users receive more comments and likes to their profile photos (Eftekhar et al., 2014). A more recent study by Dhir et al. (2016) on adolescent Facebook users indicates that extroverts tend to be more experienced in taking and sharing their photos. Furthermore, they also practice stricter protection of their shared photos on the platform (Dhir et al., 2016).

The study by Barker and Ota (2011) reveals major differences between photo sharing activity of Japanese and American females SNS users. For instance, American users engage in photo sharing activity on Facebook to connect with their peers. American Facebook users posted substantially more personal and friends' photos, meanwhile, the Japanese SNS users, all of whom were using

Mixi diary, posted far more photos of cartoon characters (Barker & Ota, 2011). Finally, a study by Liu and Fan (2015) also identified photo sharing differences among American and Chinese SNS users. Their results indicate that compared to Chinese SNS users, Americans shared more photos on SNS and also revealed more information about themselves as well as of their surroundings. Chinese users were unwilling to share photos of their family and romantic relationships on SNS. Furthermore, Chinese users also applied more privacy control mechanisms over their shared photos (Liu & Fan, 2015).

### 2.2.3 Photo tagging on Facebook

Photo tagging is one of the popular features on Facebook (Besmer et al., 2010; Dhir et al., 2015). In the context of Facebook, tagging photos can be defined as "labeling people in the photos by appending their names to the photo." However in practice, tagging is not limited only to the people present in the photos. As the tagging feature serves other purposes such as notifying and informing other people about an activity/event, people not present in the photo might also be tagged. After tagging a photo, a notification is sent to all the tagged users. The tagged user's profile name appears when someone hovers over the photo. Tagging photos on Facebook is motivating for many users as it increases the visibility of shared photos and the possibility of further sharing (Besmer et al., 2010). Furthermore, users tag their photos to fulfill various gratifications including entertainment, communicating, and appreciation (Park, Jin, & Annie Jin, 2011; Vasalou, Joinson, & Courvoisier, 2010). A recent study on photo tagging gratifications by Dhir et al. (2015) suggest that getting likes and comments, social influence, peer pressure, popularity, and affection are some of the main motivations behind this activity. Photo tagging on Facebook is the only activity on the platform that displays significant association with having more core ties (Hampton, Goulet, Marlow, & Rainie, 2012). Furthermore, positive association is also established between the length of Facebook use and tagging friends in shared photos (Hampton et al., 2012).

Finally, recent literature also indicates various factors that can impact photo tagging activity such as privacy concerns, general SNS etiquette and the likelihood of possible conflicts between the tagger and tagged person (Besmer et al., 2010; Stutzman, Gross, & Acquisti, 2013). To manage these privacy risks and possible conflicts, users tend to adopt various strategies such as requesting that the tagger untag/remove the photo, face to face confrontation, or untagging their profile manually (Lang & Barton, 2015; Rui & Stefanone, 2013; Young & Quan-Haase, 2013).

## 2.3  Privacy on SNS

The popularity of SNS has not only seen a surge of private and sensitive information but has also raised various issues concerning privacy related to these self-disclosures. Since the inception of SNS, there has been an increased interest in understanding various aspects of privacy and their impacts in reference to self-disclosures on SNS - notably Facebook. The enormous popularity and acceptance of SNS to fulfill a number of gratifications has led to an increasing amount of information and content revealed by its users (Stutzman et al., 2013; Young & Quan-Haase, 2013; Young & Quan-Haase, 2009). By sharing extensive amounts of information and content on SNS, users expose themselves as well as their content to various privacy and security threats that can lead to unintended consequences (Chang & Heo, 2014; Orito et al., 2014).

Even though SNS users report concerns with respect to their own as well as shared content privacy, these concerns are not demonstrated through their actions. This disconnect between reported privacy needs/concerns and disclosure practices is commonly referred to as "privacy paradox" (Barnes, 2006; Debatin et al., 2009). More specifically, SNS users indicate that they would like to maintain their online and personal privacy and exert control over it, but in reality, they disclose extraordinary amounts of personal and sensitive information on these platforms (Stutzman et al., 2013; Young & Quan-Haase, 2013). Moreover, many SNS users are unaware or fail to employ appropriately the protective features that are offered by most SNS (Acquisti & Gross, 2006). The privacy paradox has been a highly active research debate, yet this phenomenon has not been fully comprehended until now. SNS users tend to perform a privacy calculus i.e. calculating the expected loss of privacy and the potential benefits associated with disclosures on the platform. The actual behavior of SNS users is determined by the outcome of this privacy trade-off (Xu, Dinev, Smith, & Hart, 2011). Due to the high integration of SNS in daily lives, peer-pressure, the number of gratifications fulfilled by these platforms, and the perceived benefits of participating and disclosures mostly outweigh the expected privacy risks (Debatin et al., 2009). To understand the complex nature of the privacy paradox, prior literature has addressed various aspects, such as privacy awareness, privacy concerns, and privacy-seeking measures. These issues are some of the potential aspects of privacy that influence users' trust, actual usage, and intentions to interact with content on SNS platforms (Shin, 2010).

Digital photos on SNS are also associated with the significantly high potential for revealing sensitive and confidential information that can expose the users to various privacy threats and social embarrassment. For instance, it is highly likely that photos shared on an SNS platform can be accessed and viewed by unintended audiences (Taddicken, 2014). Unintended access and viewing of shared photos potentially increase the threats associated with misuse. Despite being aware and concerned about privacy-related issues, many SNS users continue to share increasing numbers of photos on these platforms (Litt & Hargittai, 2014; Taddicken, 2014). Recent studies provide a detailed insight into various privacy protection strategies and measures that SNS users adopt for their content especially photos (Lang & Barton, 2015; Madden, 2012; Young & Quan-Haase, 2013). For instance, frequently changing privacy settings, deleting photos regularly, controlling network strength, faking profile information, sharing photos with limited network, untagging photos, unfriending, and using photo reporting tools are some of the strategies to minimize threats and risks associated with digital photos on SNS (Lang & Barton, 2015; Pempek, Yermolayeva, & Calvert, 2009; Stutzman & Kramer-Duffield, 2010; Young & Quan-Haase, 2013).

Despite having a multitude of options to mitigate privacy-related risks associated with digital photos, many users seem to have a limited ability and in some cases, they are even unconcerned about the underlying threats (Strater & Lipford, 2008). Furthermore, in a broader context, many SNS users also lack awareness with respect to their own privacy rights, SNS privacy policies, and how their personal content is being gathered, used, and spread across SNS platforms (Molluzzo, Lawler, & Doshi, 2012).

## 2.4   Age differences in SNS behaviors

Until now, behavioral research on SNS has predominantly focused on studying young college/university students (Lee & Ma, 2012; Madden et al., 2013; Mendelson & Papacharissi, 2010; Park & Lee, 2014; Young & Quan-Haase, 2009). A limited number of studies have concentrated specifically on older SNS users as well as understanding the major differences between the age groups in SNS adoption, usage, and behaviors (Brandtzæg et al., 2010; Pfeil, Arjan, & Zaphiris, 2009). Studying young SNS users is important as they are dominant and highly active on these platforms. At the same time, it is important to cater other SNS populations as the demographics of online users is evolving rapidly. For instance, around one-third of the American adults aged 65 or above are actively using SNS (Zickuhr & Madden, 2012). Concentrating too much on young college and university students might lead to biased assumptions that cannot be generalized to other SNS users.

Recent studies indicate that young SNS users are highly involved and actively participate in various digital photos activities (Hollenbaugh & Ferris, 2014; Muscanell & Guadagno, 2012). Similarly, younger SNS users are more active and share a higher number of photos on SNS (Stefanone & Lackaff, 2009). The network size of SNS users is also reported to have a significant relationship with a number of photos shared on SNS (Stefanone & Lackaff, 2009). Likewise, young users also spend more time on various activities, have more frequent visits to the platforms and have a larger friend network (Brandtzæg et al., 2010; Muscanell & Guadagno, 2012). High involvement in various activities on these platforms also leads younger users to disclose more information such as profile photos, full names, relationship status, email addresses, school names, birth date, and current city (Taddicken, 2014; Vanderhoven, Schellens, Valcke, & Raes, 2014). Adolescents and college SNS users invest time and effort into displaying profile photos where they appear physically attractive (Siibak, 2009; Strano, 2008).

Furthermore, the motivations of using SNS varies among different age groups. For instance, the prime motivation for older SNS users is to connect with their family, grandchildren, and close acquaintances, due to which they have a small friends network on SNS (Bell et al., 2013; Brandtzæg et al., 2010). Meanwhile, the younger users engage with SNS to fulfill habitual pastimes (Papacharissi & Mendelson, 2010). Due to the varying motivations of using SNS, younger and older users engage with different features and share different types of information on the platform. Older users mostly prefer one to one communication, meanwhile, younger users are mostly interested in wider communication and engage with various forms of content including photos, videos, and comments (Hope et al., 2014; Pfeil et al., 2009). As younger users are involved in more self-disclosure, they are considered to be more vulnerable to various privacy and security risks such as stalking and cyber bullying (Brandtzæg et al., 2010; Christofides et al., 2012). Older users are also vulnerable to privacy and security risks as generally they have limited experience and skills in managing the privacy settings (Brandtzæg et al., 2010; Zickuhr & Madden, 2012). The literature on SNS privacy indicates that younger users seem less concerned about possible privacy threats due to peer-pressure and online bonding (Hoy & Milne, 2010; Litt, 2013; Stutzman & Kramer-Duffield, 2010). Meanwhile, despite having a limited friend network and activity, older users are more concerned about privacy (Hoofnagle, King, Li, & Turow, 2010).

Furthermore, age is also considered to be one of the most significant predictors of Facebook use. For instance, younger Facebook users indulge with the platform to pass time, meet new people, and also to maintain existing relationships more than the older Facebook users (Sheldon, 2008). Younger Facebook users are likely to indulge in playing games on the network to pass time, meanwhile reciprocating is the main motivation for older users (Wohn & Lee, 2013).

## 2.5   Gender differences in SNS behaviors

In addition to age, gender is also the most commonly reported variable that impacts SNS behaviors and gratifications. Similar to other channels of computer-mediated communication, usage and behavioral differences also exist among male and female SNS users. A study by Joinson (2008) found that gender significantly predicts Facebook usage frequency, as females visit Facebook more frequently than males (Joinson, 2008). A similar finding has also been reported by Sheldon (2008). Male Facebook users are more likely to access Facebook for meeting new people and developing new relationships, meanwhile, female users visit Facebook for passing time, profile surveillance, maintain existing relationships, interpersonal communication, and entertainment purposes (Hunt et al., 2012; Sheldon, 2008). Moreover, women have more Facebook friends and express higher levels of satisfaction with Facebook (Sheldon, 2008). On features usage, males are found to be associated with higher usage of Facebook chat features than females (Smock et al., 2011). Similarly, female users play games on Facebook to cope and are also more active in gift exchanges and customizing the space than men (Wohn & Lee, 2013).

Female users are more likely to be a member of SNS, are more active in socializing activities on SNS, and have a larger friends network (McAndrew & Jeong, 2012; Pfeil et al., 2009). Compared to males, female SNS users communicate more, receive more comments to their posted content, and are also more actively followed by other network members (Pfeil et al., 2009). The results of these studies clearly indicate that the behaviors and motivations of the gender groups have fundamental differences.

As both gender groups display varying usage motivations, their visuals-related activities and behaviors on SNS also vary considerably. In general, compared to males, female SNS users are more active on SNS platforms as well as photos related activities (Mendelson & Papacharissi, 2010; Pempek et al., 2009; Stefanone & Lackaff, 2009). Females SNS users spend more time in managing their profiles as well as sharing photos (Stefanone et al., 2010). Starno's (2009) study of 18-24 years old students found that females change their SNS profile photos more often than men do. Moreover, females carefully choose their profile photos to depict friendship and also portray their attractiveness (Strano, 2008). Similarly, females SNS users are more likely to tag themselves and others in their photos (Mendelson & Papacharissi, 2010). Photos of family, friends, and vacations are shared by female Facebook users mostly with a selected audience (Litt & Hargittai, 2014; McAndrew & Jeong, 2012). On the other hand, males tend to share sports-related photos and videos publicly (Litt & Hargittai, 2014; McAndrew & Jeong, 2012). Females prefer to share photos with restricted access as they are comparatively more concerned about their privacy and are more likely to be victims of stalking, online harassment, and cyber bullying than males (Fogel & Nehmad, 2009; Hoy & Milne, 2010; Li-Barber, 2012).

Due to higher concerns over the privacy of their photos, women tend to be more cautious and actively seek various privacy-protecting measures to mitigate these concerns. For instance, many female SNS users conceal their real identity, frequently review their privacy settings, frequently delete their photos, and untag themselves from unwanted photos (boyd & Hargittai, 2010; Litt, 2013; McAndrew & Jeong, 2012). Moreover, they have restrictive personal profiles and engage actively in unfriending activity (Madden, 2012). Similarly, females SNS users are also careful in sharing their personal details such as phone number and address (Fogel & Nehmad, 2009). Interestingly, despite having higher privacy concerns, females users are comparatively more satisfied with SNS and their engagement and disclosures (e.g. photo sharing and photo tagging) are higher than their counterparts (Kumar & Schoenebeck, 2015; Li-Barber, 2012).

With respect to gender attitudes towards Facebook photo tagging, women are more likely to tag their photos as well as get tagged by others (Mendelson & Papacharissi, 2010). As females are more often tagged by others, they also engage more in untagging themselves than male Facebook users (Pempek et al., 2009). Privacy concerns and displeasure with their appearance are some of the main reasons for disassociating themselves from tagged photos (Pempek et al., 2009; Pesce et al., 2012).

# 3. AN OVERVIEW OF THE ORIGINAL ARTICLES

The main aim of this dissertation is to understand the current practices associated with digital photos in computer-mediated platforms and examine the factors associated with photo sharing and photo tagging practices. The dissertation comprises five empirical studies that focus on understanding these practices and associated relationships. Study I investigate the factors associated with tagging photos in personal photo collections. Study II explores the factors that influence photo tagging activity of Facebook users. Study III identifies various factors that motivate photo-sharing activity on Facebook. Study IV models the impact of various privacy aspects, trust, and user activity on user's intentions to share photos on Facebook. Finally, study V investigates the impact of gender and age groups on various aspects of privacy, trust, and intentions in relation to Facebook photo sharing activity. In this chapter, a brief overview of all five empirical studies is presented.

## 3.1 Study I

### 3.1.1 Aims

The main purpose of study I was to explore the current practices surrounding personal photos with a focus on the tagging of personal photos. Furthermore, it aimed to obtain insights into various factors that limit the usage of tagging features and how tagging practice can be made more acceptable to the users. In the context of tagging photos in personal repositories (e.g. photos saved on personal computer/drives) most of the relevant research was carried out in the latter half of the previous decade, hence, it is important to review whether tagging practices surrounding photos have changed over time or not. It is also important to study photo tagging in this context as in recent years photo management tools have become highly advanced as well as the number of photos people capture have also increased substantially. Furthermore, with the ever-increasing popularity of tagging feature in a number of online platforms, it is important to get insights from the perspective of tagging in personal repositories for comparison purposes. To answer these questions, a user study on a prototype photo tagging application "SmartImages" was carried out. The application allows users to view, browse, search, and tag their personal photos. The application also includes novel features such as "tag suggestions", "adding tags", and "similar photo search."

### 3.1.2 Participants and procedure

In total, 15 participants were recruited for this user study. The study was carried out in the usability lab of the Department of Computer Science, Aalto University. Ten participants were males while the remaining five were females. Nine participants were 18-24 years old, five were aged 25-34 years, and one participant was 39 years old.

The user study was composed of a pre-task questionnaire, task analysis, and a post-task questionnaire. In the pre-task questionnaire, all the participants were requested to provide relevant demographic details. The task analysis method, which is one of the commonly used User Centered Design (UCD) techniques, was used for gaining insight into current tagging practices and related user needs (Hackos & Redish, 1998). A set of scenarios were developed in relation to photo tagging with an aim to gain a deeper understanding of the actual goals of photo tagging and how the users achieved them by completing the tasks. In total, there were nine main tasks addressing navigation, adding/removing tags, keywords, and face tagging features of the applications. After performing each task, open-ended questions were carried out to discuss the task related problems and design issues. The usefulness of the specific feature in real life situations and alternate solutions were also discussed. After completing all the test tasks, in the post-task phase discussion about the tested application, its features, and new ideas was carried out. All the sessions were video-recorded with the prior permission of the participants. The author later manually transcribed all the recorded sessions for further analysis.

### 3.1.3 Measures

**Demographics:** All the study participants were requested to provide their demographic details, including gender (evaluated as Male = 1, Female = 2) and age (assessed using three age groups).

**Digital photos practices:** Before initiating the task analysis phase, all the study participants went through an open-ended pre-task questionnaire to gain an understanding of their photo related practices on various computer-mediated platforms. More specifically, participants were asked multiple questions in the context of managing, searching, tagging, and sharing photos.

**Task analysis:** In this phase the users were requested to perform various tasks on the SmartImages application. The main aim of the tasks was to evaluate a set of key activities including navigation of the application, adding and removing photo tags, keywords and similarity search, and face tagging. During the phase, the users were requested to think aloud while performing the tasks.

### 3.1.4 Analyses

To organize the data gathered during the sessions, an affinity diagramming technique was utilized (Hackos & Redish, 1998). All the responses were grouped by each task and sub-task. Open-ended answers were also grouped into meaningful themes to get a thematic understanding of the topic. Based on the consolidated data, main themes were outlined. The themes that emerged from the consolidated data were general tagging practices, real-life usage and tagging needs, UI related findings, and new ideas.

### 3.1.5 Results

Results from this study suggest that mobile handset is the most preferred device for capturing photos. Four users captured photos from their mobile phones on a daily basis, nine engaged in the activity at least once a week, while the remaining two took photos less frequently. Tablets are also used for capturing photos but less often. Digital cameras are also used less frequently as they are meant for taking photos mostly on scheduled and special events. The number of photos stored by study respondents on various devices (mobile phone, tablet device, PC, laptop, cloud services) ranged from 1000 to 10,000. Respondents also stated that they rarely engaged in photo organization activity, as only one participant was actively involved in managing them on a regular basis. Even though all the study participants knew about photo tagging, within the context of personal photo management the activity is rarely practiced.

### 3.1.6 Discussion

The study results indicate that most of the users engage actively in photo-related practices, especially capturing and sharing them online. This finding is consistent with prior literature pointing to the popularity and adoption of photos on the internet (Besmer et al., 2010; Litt & Hargittai, 2014). With the in-creasing amount of photos, multiple capturing and sharing devices, and online/offline storage for managing photos are becoming increasingly difficult. The results also suggest that mobile phones are have become the primary photo capturing and sharing device. Increasingly, fewer people are using digital cameras for capturing everyday photos. Digital cameras are mostly used for special occasions as they are considered a device for serious photography.

For locating photos, manual browsing and memory is used. Search features are rarely used for locating photos as the participants do not rename their photos and most of the photos have default naming. Renaming photos is considered a daunting task and rarely practiced by the participants. Organization of photos is mostly limited to creating new folders, naming folders, moving photos from one folder to another, and deleting multiple copies.

The results also suggest that almost all the participants are aware of the term "tagging", though it is mostly associated with the Facebook tagging feature. In the context of personal photo collections, the actual practice of tagging photos is uncommon. Tagging photos in personal repositories is considered to be an extra task in the photo-work process. Even though many participants are aware of the advantages of photo tagging, they are hesitant to engage in the activity. The time and effort required to tag photos, laziness, and the thinking process to come up with good tags were the main reasons for not indulging in the practice. Moreover, the limited visibility of tagging features, limited interactivity and playfulness, and non-existent social features further add to the non-acceptance of photo tagging practice.

The users expect that automation, high accuracy, and relevance of the tagging results can be important factors in the adoption and acceptance of tagging feature. The respondents also indicated that new tagging solutions should require minimal effort, automated face and objects tagging, and most importantly, incorporate social elements for user acceptance.

## 3.2 Study II

### 3.2.1 Aims

The purpose of study II was to examine the practices associated with photo tagging activity on Facebook among teenagers in India. This study also explored the factors associated with perceived usefulness, liking, and disliking of the photo tagging practice. Furthermore, how photo tagging uses and gratifications, liking, and disliking vary among gender groups in India is also explored.

### 3.2.2 Participants and procedure

This qualitative study was based on two sections. The first section of the study was composed of basic details of the study participants including gender, age, and ownership of mobile phone and digital camera. The second section consisted eight open-ended questions on Facebook photo tagging practices. In this study, 67 school children aged 13-18 years participated. In total, 35 female and 32 male participants completed the two-part qualitative study. Ten students were aged 13 years, 25 were 14 years old, and 15 were 15 years old. 11 teenagers were aged 16 while the remaining 6 were 17-18 years old. 29 participants owned a mobile phone while 27 of them owned a digital camera. Despite the fact that less than half owned a mobile handset and digital camera, all the respondents mentioned that they have regular access to these devices through their siblings or parents.

The study was developed in English and organized at two English-medium schools in Punjab, India in December 2013. A pilot study was carried out before the actual studies. Six teenage students participated in the pilot session to assess the average time to complete the study, language, and terminology understanding. Based on the feedback from the pilot study, necessary amendments were made to the study. Before initiating each study, an information session was conducted to brief participants about the study and related practicalities. During each session, one author along with a school teacher supervised the 40-minute sessions.

### 3.2.3 Measures

Eight open-ended questions were used to get input from the study participants. These questions included motivations for practicing photo tagging, perceived usefulness of photo tagging practice, likes and dislikes about photo tagging, attitude when others tag them in digital photos, and type of photos one likes or dislikes to get tagged.

### 3.2.4 Analyses

After gathering the qualitative data, the researchers processed the data by making necessary corrections (spelling mistakes, expanding abbreviations for clarity, etc.) and created an order for further analysis. Using the affinity diagramming technique, all the processed data was analyzed. The data was grouped into meaningful themes by two authors independently. Finally, both the researchers compared and discussed the grouped themes to form a collective result.

### 3.2.5 Results

Compared to girls, teenage boys in India are far more engaged in the Facebook tagging practice. More than half of the boys are highly active in tagging practice while very few girls mentioned their active involvement with the practice. Two-thirds of the teenage girls do not engage in the practice at all. Furthermore, the responses of the teenage boys were thorough and explained in detail, even with accompanying examples in some cases. On the other hand, the answers provided by girls were pretty concise and very short such as "yes", "no", "do not like".

### 3.2.6 Discussion

The results indicate a clear divide among gender groups in Indian teenagers with respect to photo tagging practices and attitudes. Photo tagging activity is highly embraced and liked by a majority of the Indian teenage boys. On the other hand, Indian teenage girls minimally engage with tagging practices and in general display a disassociation with and dislike of the practice.

Teenage boys considered tagging as a social symbol and a tool for getting likes and comments on their shared photos. Getting more likes and comments lead them to gain status, attention, and visibility within their peer groups. It is interesting to note personal appearance in photos, which is generally an important consideration for girls, is highly valued among teenage boys in India. Teenage boys prefer to be tagged and share photos in which they appear handsome and cool. They would prefer to untag photos where their appearance is not attractive. On the other hand, for girls, personal appearance is not an important concern as they avoid photo related practices in the first place. Some of the teenage girls who mentioned their involvement in tagging practice on Facebook engage with it mainly for fun and self-expression purposes.

The majority of the boys consider Facebook photo tagging to be a useful activity as they can fulfill various gratifications including social status, prestige, and self-expression. Meanwhile, the majority of girls were unfavorable to tagging, as they perceived the activity to be useless. The main reasons behind this perception were attributed towards time wastage and the non-productive nature of the activity. Furthermore, parental influence and cultural norms might have a strong association with this negative perception among teenage girls. Teenage girls also expressed a high level of privacy concerns and intrusion, which can also be one of the main reasons behind this negative perception. This finding strongly aligns with prior studies on SNS indicating higher priva-cy concerns among female SNS users (Pempek et al., 2009; Rui & Stefanone, 2013).

## 3.3 Study III

### 3.3.1 Aims

Photo sharing has been recognized as the most popular activity on Facebook yet the research examining the photo sharing practices on the platforms is highly limited (Eftekhar et al., 2014). The main aim of this study was to fill this gap by assessing the uses and gratifications of Facebook photo sharing. In addition to this, the study also examined how gender, age, and number of photos shared relate to U&G of Facebook photo sharing. A theoretical framework based on U&G was utilized to address the different research questions of this study.

### 3.3.2 Participants and procedure

A cross-sectional online survey was developed and hosted during the first quarter of 2014 for one month. Any Facebook user over 18 years engaged in sharing photos on Facebook was eligible to participate in the study. The survey was not targeted towards any specific demographic group(s). By browsing through the "suggested groups" list displayed on the Facebook profile of the first author, a list of potential groups was created. During this process, every effort was exhausted to ensure that the selected groups were of general public interest and did not target photo-savvy users in any way. After the initial selection of the potential groups, the list was refined once again to meet the criteria mentioned above. Once the final selection was made, we sent personal chat messages to group administrators together with a brief about the survey and the request to post the survey link on their group page.

In total, the online survey contained 35 questions. During one month 442 respondents completed the survey. Before starting the data analysis, we removed the incomplete responses as well as the respondents who do not use Facebook for sharing their photos. The data cleaning process yielded 368 valid responses that were utilized for data analysis.

The male and female ratio was quite balanced as 194 (52.7%) were females, while remaining 174 (47.3%) were males. 53 (14.4%) respondents were 18-25 years old, 146 (39.6%) were 26-35 years old while 96 participants (26.1%) were aged between 36-45 years. 65 responses (17.7%) came from 46-60 year olds, meanwhile, the rest of the 8 participants (2.1%) were over 60 years old.

### 3.3.3  Measures

**Photo sharing U&G:** After an extensive review of previous literature concentrating on U&G of SNS and photo sharing studies on various computer-mediated platforms, we devised a 26-item instrument. This scale addresses eight photo sharing gratifications on Facebook including affection, attention seeking, disclosure, entertainment, habitual pastime, information sharing, social influence, and social interaction. A five-point Likert-scale ranging from 1 (strongly disagree) to 5 (strongly agree) was used to assess this scale.

**Demographics:** All the study participants were requested to provide their demographic details, including gender (evaluated as Male = 1, Female = 2) and age group (assessed using five age groups).

### 3.3.4  Analyses

SPSS 21.0 was utilized during the data cleaning and analysis process. Before the actual data analysis of 442 total responses, we carried out a data cleaning process by utilizing missing value analysis (MVA) feature. In total, 40 cases were deleted, as they were not using Facebook for sharing their photos. Furthermore, 34 cases infected with over 50% of the missing values were also deleted during the process. After the data cleaning process, a data set of 368 respondents was utilized for detailed data analysis. To examine the gratifications of Facebook photo sharing we utilized exploratory factor analysis (EFA) with Maximum Likelihood Estimation (MLE) algorithm with "Varimax rotation" on the cleaned empirical dataset. All the survey items that did not fulfill the threshold limit of .50 factor-loading were deleted from item pool. To understand the relationship of age with resulting gratifications, correlation analysis was carried out. Similarly, correlation analysis was conducted to access the relationship between the number of photos shared on Facebook and resulting gratifications. Furthermore, an independent samples t-test was conducted to understand the gender differences among different photo sharing U&G.

### 3.3.5  Results

The EFA process yielded six gratifications of Facebook photo sharing, namely Affection (AF), Attention seeking (AS), Disclosure (DE), Habitual pastime (HP), Information sharing (IS) and Social influence (SI). The total variance explained by EFA resulted in 77.68%, which is considered excellent. Furthermore, all the resulted gratifications successfully fulfilled the Kaiser criterion, i.e., eigenvalue > 1.0. Results from the correlation analysis indicate that age has a significant positive relationship with social influence (r = .15, N = 368) and disclosure (r = .15, N = 368) gratifications. These results indicate that photo sharing activity is used more to fulfill disclosure and social influence by the older Facebook users. The gender differences were also revealed between two photo sharing U&G, namely habit and disclosure. The results from the independent samples t-test suggest that, compared to females, males actively seek higher habit (t = 3.37, p < .01, Mean = 2.56, SD = .89 vs. Mean = 2.25, SD = .87) and disclosure (t = 3.10, p < .01, Mean = 2.97, SD = .83 vs. Mean = 2.71, SD = .74) U&G from Facebook photo sharing.

   The results also reveal that the number of photos shared on Facebook negatively correlates with habit (r = -.41, N = 368) and information sharing (r = -.19, N = 368) gratifications. Finally, results from the Pearson correlation analysis suggest that habit positively correlates with disclosure (r =. 23), attention seeking (r = .24), social influence (r = .25), and information sharing (r = .17). Affection was in low positive correlation with disclosure (r = .14) and attention seeking (r = .15). Disclosure was in positive correlation with attention seeking (r = .34), social influence (r = .24), and information sharing (r = .14). Finally, attention seeking was in medium positive correlation with social influence (r = .36), and social influence correlated positively with information sharing gratification (r =. 19).

### 3.3.6 Discussion

This study contributes to the growing body of knowledge on user behaviors in various activities on SNS. By utilizing the U&G framework, our study identified six U&G of photo-sharing activity on Facebook. These identified U&G include affection, attention seeking, disclosure, habit, information sharing, and social influence.

   Being affectionate in the form of being thankful and considerate to others has been regarded as an important SNS gratification (Cheng et al., 2014; Quan-Haase & Young, 2010). In the case of photo sharing on Facebook, we found that seeking affection in the form of "likes" and "comments" from other users in the network is an important U&G sought by the respondents. Seeking attention from others in the social network has also been highlighted in the previous literature (Park et al., 2009; Urista, Dong, & Day, 2009). Results from this study indicate that photos are one form of content that users share with others on Facebook with an intention to gain attention. Gaining attention from other members has also been highlighted as an important motivational factor on the popular photo sharing platform, Flickr (Malinen, 2011).

Content shared by SNS users is one of the main lifelines of these platforms. Disclosure on SNS (e.g. text, photos, and location information) has been recognized as an important U&G of these platforms (Hollenbaugh & Ferris, 2014; Mendelson & Papacharissi, 2010). Even though previous literature indicates strong user concerns over privacy in SNS (Fogel & Nehmad, 2009; Orito et al., 2014), our results indicate that users share photos on Facebook to disclose more about themselves as well as people around them. Despite the tension between privacy concerns and sharing photos (and other content), our findings strongly align with previous research indicating high levels of information disclosures even having privacy concerns (Litt & Hargittai, 2014; Taddicken, 2014). Habit, which is also an important U&G of Facebook usage in general (Quan-Haase & Young, 2010; Smock et al., 2011), and online photo sharing (Ames et al., 2010; Miller & Edwards, 2007) is also validated in our study as an important gratification of Facebook photo sharing activity.

Results also found out that Facebook users share photos for information sharing purposes. These results confirm the prior findings of sharing photos on the web for informational purposes (Goh et al., 2009; Naaman et al., 2008). Results from our study also endorse that Facebook users share photos to seek social influence gratification. To follow a societal trend or to be a part of a peer-group has also been highlighted as an important gratification in social media usage (Papacharissi & Rubin, 2000) as well as online photo sharing activity (Oeldorf-Hirsch & Sundar, 2010).

Our study also reveals interesting relationships between age, gender, and number of photos shared with photo sharing gratifications. For instance, compared to young, older users seek more disclosure and social influence gratifications from sharing photos on Facebook. As younger users are generally more aware of various aspects of SNS usage (Brandtzæg et al., 2010; Zickuhr & Madden, 2012), they might be more aware of what to disclose on the platform. Similarly, it is likely that younger Facebook users might have more privacy concerns that lead them to be more cautious and restrict their disclosures. SNS being a new phenomenon for older Facebook users could be one of the main explanations behind the higher level of social influence gratification among them. For older groups, photo sharing on Facebook might be a cool new trend to follow that their children and other relatives have been engaged with for a while. For the older Facebook users, engagement with photo sharing activity might also help them to be an active part of society.

Finally, our study results indicate that compared to women, men sought more disclosure and habit gratifications from photo sharing activity. Hence, it can be argued that men use Facebook photo sharing activity to disclose more about themselves as well as others around them. This finding is also supported by prior literature that men are less concerned about their online privacy and disclose more personal information (Fogel & Nehmad, 2009; Li-Barber, 2012).

## 3.4 Study IV

### 3.4.1 Aims

The main aim of this study is to understand the impact of various privacy issues (e.g. privacy concerns, privacy awareness, and privacy-seeking behavior), trust and activity on users' intentions to share photos on Facebook. Even though photo sharing activity is practiced daily by many Facebook users, a limited amount of research has been carried out to understand the privacy-related attitudes and relevant aspects in the context of Facebook photo sharing (Litt & Hargittai, 2014). To address this research gap, this study investigates the impact of privacy awareness, privacy-seeking behavior and privacy concerns on shaping an individual's trust and activity level. Furthermore, the study also examines the mediating role of trust and activity levels on intentions to share photos on Facebook.

### 3.4.2 Participants and procedure

The data was collected using an online survey hosted on a web-based survey platform from December 2014 until January 2015. The target audience was Facebook users over 18 years old engaged in sharing their photos on the platform. The survey was publicized on multiple Facebook groups by employing the snowball sampling technique. We encouraged all our participants to forward the survey details to their friends and acquaintances. Participation was completely voluntary and anonymous. During the preliminary data screening respondents not using Facebook, incomplete responses, and obvious outliers were removed that yielded 378 (n=378) valid responses. Male-female ratio was pretty balanced as 49.7% (n=188) were females, and 50.3% (n=190) were male participants. 11% (n=43) of the study participants were aged 18-24 years, 39% (n=146) were 25-34 years old, 29% (n=108) were 35-44 years old, 13% (n=50) were 45-54 years old, while the remaining 8% (n=31) were over 55 years old.

### 3.4.3 Measures

**Demographics:** All the respondents were requested to provide details about gender, age, educational background, and country of residence.

**Facebook and photo sharing activity:** Four items were included in the survey to get a deeper understanding of respondents' usage of Facebook and photos. Tenure on Facebook was measured on a five-point Likert scale (1=less than 6 months, 2=6-12 months, 3=1-3 years, 4=3-5 years, 5=over 5 years), frequency of visiting Facebook profile was also measured on a five-point Likert scale (1=less than once each week, 2=once each week, 3=several times each week, 4=once each day, 5=several times each day). The number of photos shared online and number of photos shared on Facebook was measured on a 4 point Likert scale (1=less than 10, 2=between 10 and 20, 3=between 20 and 50, 4=more than 50).

**Privacy, Trust, and Sharing Intentions:** To measure privacy (privacy concerns, privacy-seeking behavior, and privacy awareness), trust, and sharing intention, 23 statements were measured on a five-point Likert scale (1 = "strongly disagree" and 5 = "strongly agree").

### 3.4.4 Analyses

The data was analyzed using a partial least squares path modeling (PLS) approach. PLS has been increasingly applied as a tool for theory testing in the IS community and it is currently one of the most common quantitative data analysis methods used (Gerow, Grover, Roberts, & Thatcher, 2010). Before initiating the actual analysis, the measurement and structural models were evaluated. The measurement model was evaluated by ensuring the reliability of the construct using Cronbach's alpha and Composite Reliability (CR), average variance extracted (AVE), and discriminant validity of the latent variables. Later, the structural model was validated to confirm causal relationships and consistency of the data.

### 3.4.5 Results

Outer loadings of all the indicators were higher than 0.7, hence satisfying the requirement for indicator reliability. Similarly, all of the constructs satisfied the reliability requirements CR > .70 for exploratory research. Moreover, all of the constructs have an AVE > 0.5, which satisfies the requirements for convergent validity (Hair,J.F., Hult, G.T.M., Ringle, C.M., & Sarstedt, 2014). The discriminant validity is satisfied by comparing the square root of the AVE of each construct with the correlations between constructs. Next, the structural model was evaluated by examining the values of R2, Q2 test for predictive relevance, the size of the coefficients of paths, and the stability of the estimations by means of the t-statistics obtained in the bootstrap with 3000 samples. Table 6 presents a summary of all the proposed hypotheses, and whether they were accepted or rejected by the work.

**Table 6.** Summary of study hypothesis

| Study hypothesis | Hypothesis |
|---|---|
| Privacy awareness → Trust | Accepted |
| Privacy awareness → Activity | Accepted |
| Privacy seeking behavior → Trust | *Rejected* |
| Privacy seeking behavior → Activity | Accepted |
| Privacy concerns → Trust | *Rejected* |
| Privacy concerns → Activity | *Rejected* |
| Trust → Sharing intentions | Accepted |
| Activity → Sharing intentions | Accepted |

### 3.4.6  Discussion

Privacy awareness has the strongest relationship with trust, suggesting that Facebook users having high levels of privacy awareness display higher trust levels with Facebook. This finding is in line with prior IS literature indicating that users with higher privacy awareness exhibit higher levels of trust with the service as they are more knowledgeable about the reality and actual expectations in the context of their privacy (Hoadley, Xu, Lee, & Rosson, 2010). Higher awareness about privacy can lead the Facebook users to more conscious actions when sharing their photos on the platform. For instance, it is quite likely that users who are more privacy aware selectively and consciously share their photos with others on the platform. Being cautious and selective makes them confident enough that their shared photos won't be misused on the network, which ultimately leads to higher level of trust on Facebook.

The study results also indicate a positive relationship between privacy awareness and Facebook activity. This relationship suggests that users with high levels of privacy awareness are more likely to have a high level of activity on Facebook. It can be concluded that users with higher privacy awareness are more realistic and know more about necessary measures to protect their privacy. One of the important privacy measures, sharing with a limited network of friends (Litt & Hargittai, 2014; Stutzman et al., 2013), might encourage them to share more photos within the network as they feel confident that the close network will not violate their privacy. Hence, it can be deduced that being privacy aware leads to higher trust and activity on Facebook.

Similarly, privacy-seeking behavior also significantly impact Facebook activity, though it displays limited impact on trust level with Facebook. This relationship suggests that users who actively engage in various strategies to protect themselves feel more confident, leading them to visit Facebook more frequently and share more content. This finding can be mirrored to the earlier findings suggesting that Facebook users who spend more time on the site also have higher levels of confidence in using Facebook's privacy settings (boyd & Hargittai, 2010). It is also likely that individuals who use Facebook more often are also disclosing more information so they have more information to protect, and thus engage in a wider array of privacy practices. Engaging more with various privacy protection strategies minimizes privacy risks that ultimately lead the user to share more photos on the platform with confidence.

Interestingly, our results indicate a non-significant relationship between privacy concerns – trust and privacy concerns – user activity. These findings suggest that privacy concerns have no impact on levels of trust and activity of the users. Previous research within SNS domains also reveals high levels of privacy concerns among SNS users, even though these privacy concerns do not translate effectively to their actual behaviors (Debatin et al., 2009; Taddicken, 2014). It is highly likely that, despite having privacy concerns, users consider Facebook to be a significant platform for sharing their pictures with others. Having privacy concerns might have led many Facebook users to engage in privacy protection measures leading them to be confident enough to keep sharing their photos with others. This paradox of privacy concerns and self-disclosure has also been highlighted in previous literature (boyd & Hargittai, 2010; Taddicken, 2014).

Finally, trust and user activity display a significantly positive relationship with users' intentions to share photos on the platform. This result indicates that trust in Facebook encourages users to keep sharing their photos in the future. This finding also aligns with the previous studies indicating trust as one of the vital aspects influencing users' activity and their willingness to disclose and share content on Facebook (Dwyer, Hiltz, & Passerini, 2007). Further-more, the likelihood of photo sharing is also higher if the users are already active on Facebook. Previous research on Facebook also found that engagement levels with Facebook positively correlate with information disclosure on the platform (Chang & Heo, 2014).

## 3.5 Study V

### 3.5.1 Aims

The aim of this study was to investigate the similarities and differences in privacy attitudes and trust among age and gender groups with respect to photo sharing practices on Facebook. A significant amount of literature on SNS behaviors and privacy has concentrated on young adults, notably college/university students (Blank, Bolsover, & Dubois, 2014; Madden et al., 2013; Vanderhoven et al., 2014). As the prior research indicates varying motivations for Facebook usage among young and old users, it is quite plausible that their privacy concerns, attitudes, awareness, and trust may also vary regarding sharing photos on Facebook. Thus, this paper investigates whether different age groups vary in privacy concerns, privacy awareness, privacy-seeking behaviors, trust, and activity regarding sharing photos on Facebook. Previous research also highlights usage differences and varying gratifications of Facebook among gender groups. Hence, it is likely that the various aspects of privacy and trust may also vary among males and females with respect to Facebook photo sharing activity. Thus, this study also investigates whether gender groups vary in privacy concerns, privacy awareness, privacy-seeking behaviors, trust, and activity regarding sharing photos on Facebook.

### 3.5.2 Participants and procedure

The study participants and procedure were the same as in Study IV.

### 3.5.3 Measures

The study measures were the same as those utilized in Study IV
**Demographics:** The demographic measure was the same as that in Study IV.
**Facebook and photo sharing activity:** This measure was the same as those in Study IV.
**Privacy, Trust, and Sharing Intentions:** A 23-item measure developed in Study IV was utilized.

### 3.5.4  Analyses

An independent-samples t-test was conducted to compare the effect of gender on privacy concerns, privacy awareness, privacy-seeking behavior, trust, Facebook activity, and photo sharing intentions. Furthermore, one-way analysis of variance (ANOVA) was carried out to explore the impact of age on various privacy aspects, trust, Facebook activity and photo sharing intentions.

### 3.5.5  Results

There were no significant differences between the privacy awareness, privacy-seeking, trust, activity and sharing intentions scores. However, a significant difference ($t$ (378) = 2.02, $p$ = 0.043) in scores between males (M=3.68, SD=0.92) and females (M=3.86, SD=0.79) was revealed for privacy concerns. An alpha level of .05 was used for all statistical tests.

 The ANOVA test revealed that privacy concerns, privacy-seeking behavior, Facebook activity and trust on Facebook significantly differ across the age groups. The effect sizes, calculated using eta squared, varies between 0.03 and 0.06. Furthermore, inter-correlations for the constructs were calculated. An alpha level of .05 was used for all statistical tests.

### 3.5.6  Discussion

With respect to sharing photos on Facebook, youngest respondents display highest privacy concerns and these concerns decrease with increasing age. These findings are consistent with earlier studies reporting high levels of privacy concerns among young Facebook users (Brandtzæg et al., 2010). Multiple factors such as higher activity on Facebook, high number of friends and the dispersed network scatter of young Facebook users can be attributed to this finding. On the other hand, older Facebook users' might be less concerned as their sharing activity is mostly targeted towards specific people including their children, grandchildren, and close acquaintances (Pfeil et al., 2009).

 The age differences in privacy-seeking behavior also display a pattern similar to privacy concerns. Privacy-seeking behavior to protect their photos is highest among young Facebook users and this behavior significantly decreases with increase in age. This finding is consistent with previous research on SNS indicating the active involvement of young Facebook users in untagging photos, unfriending, restricting profile access, and deleting photos regularly (Blank et al., 2014; Madden, 2012; Young & Quan-Haase, 2013). Furthermore, com-pared to older participants, the young consider Facebook to be a trustworthy platform for sharing their photos. This finding also echoes prior research on SNS indicating higher levels of trust among younger Facebook users in comparison to older users (Torres, 2012). Similarly, study results indicate that Facebook activity also negatively correlates with age. This finding is also consistent with prior literature reporting young users as more active and spending more time on the platform (McAndrew & Jeong, 2012). Despite having higher privacy concerns and higher privacy-seeking behavior, privacy awareness of younger Facebook users do not differ significantly in comparison with older groups. This finding is consistent with Hoofnagle et al.'s (2010) finding indicating no differences among various age groups with respect to privacy awareness.

The study results on gender differences are consistent with previous findings that report higher levels of privacy concerns among female Facebook users (Fogel & Nehmad, 2009). As women generally communicate more, have more friends, and are more actively followed on SNS (Pfeil et al., 2009) that can also increase their privacy concerns. It is also likely that some of the female users (or their acquaintances) have already experienced privacy threats such as stalking or cyber bullying leading to increased privacy concerns with respect to sharing photos on Facebook.

Finally, there were no significant differences among the gender groups with respect to privacy awareness, privacy-seeking behavior, trust, Facebook activity, and photo sharing intentions. Similar to previous research on SNS privacy that highlight the blurring divide between the behavior of gender groups, our results also endorse this notion (Blank et al., 2014; boyd & Hargittai, 2010).

# 4. DISCUSSION

The purpose of this dissertation was to underpin the important issues concerning digital-photo interaction in computer-mediated platforms. During the past few years, photo-sharing and photo-tagging activities have become increasingly popular and considered an important element of numerous computer-mediated platforms. However, despite this increased popularity and adoption, the understanding of the associated practices and motivations has been inadequate. Prior studies examining digital-photo interaction in computer-mediated platforms have predominantly focused on domestic photography, mobile phone photography, and the online photo sharing platform Flickr. These pioneering studies, undoubtedly provide excellent insights into various aspects of photos related activities and features. However, most of the studied platforms offer no or somewhat limited social interactivity and communicability elements. The uses and gratifications of SNS differ substantially from the previously studied platforms as social interactivity and communicability formulate the core of the majority of SNS platforms.

This dissertation bridges the aforementioned research gaps in digital-photo interaction by utilizing a mix of quantitative and qualitative methods including online cross-sectional surveys, open-ended questionnaire, and a task evaluation approach. By utilizing these methods, this dissertation provides a nuanced understanding of some aspects of users' practices, motivations and the complex tensions that exist in digital-photo interaction in the context of personal photo repositories and leading SNS platform: Facebook. Furthermore, this dissertation provides new knowledge concerning demographic attributes (age and gender), privacy, trust, and user intentions in relation to digital-photo interaction on Facebook.

This dissertation work highlight some of the important aspects associated with digital photos on computer-mediated platforms. A deeper understanding of user behaviors, motivations, and needs in the context of digital photos will contribute to improving the designs of photos sharing and tagging features as well as privacy tools for managing photos on these platforms. Based on the first two studies, it becomes evident that photo tagging behaviors, motivations, and usage differ substantially in different computer-mediated platforms. Users are more knowledgeable, interested, and willing to engage with the tagging feature on SNS as compared to personal photo repositories. Among other reasons, interactivity, sociability, and communicability features bundled with photo tagging activity on Facebook can be attributed to this preference.

The second study that was carried out with adolescent students in India reveals a substantial divide among the gender groups. Even though, prior literature indicate gender differences in SNS usage and behaviors, study II highlights novel aspects of this divide i.e. how parental and societal influences impact adoption and usage of technology in India. This influence has a clear impact on the adoption, usage, and ultimately perceptions of the photo tagging activity among teenage Indian girls. The preceding three studies that focused on Facebook photo sharing activity also revealed novel and interesting findings. For instance, Study III that identified six photo-sharing U&G, one of them "seeking affection" (in the form of likes and comments) is different from the previously identified U&G of specific SNS features. Results from these studies also indicate that even though privacy concerns are high among all the users groups, yet people continue to share photos on Facebook to fulfill various gratifications. Similarly, these studies also reveal a number of age and gender differences with respect to photo-sharing activity on Facebook. For instance, younger Facebook users are more concerned about the privacy of their photos and they are also more active than other users to protect their photos by engaging in privacy-seeking behaviors (e.g. deleting photos). Interestingly, younger Facebook users also display a higher level of trust and activity on Facebook. Finally, the results from study V also reveal that females are much more concerned about the privacy of their photos than the male counterparts.

In the next sections, various theoretical and practical implications of this dissertation are discussed and presented. In addition to this, various study limitations are also presented. The chapter concludes by presenting possible avenues for future research.

## 4.1 Theoretical implications

This dissertation offers different theoretical implications for researchers in the domain of computer-mediated communication, new media, HCI, SNS, and privacy.

First, in a wider context, this dissertation contributes to the existing body of knowledge involved in understanding the usage practices and behaviors of SNS users. Second, the study results contribute to the emerging literature that investigates feature-specific U&G of computer-mediated platforms. By specifically studying digital-photo interaction on Facebook, this dissertation provides an understanding of usage and influencing factors at a feature-specific level of a particular SNS platform. Most of the prior research on SNS has predominately focused on SNS in general, thus overlooking feature-specific usage and motivation perspective. More recently, scholars have underlined the need to study SNS platforms at a granular level by concentrating on feature-specific aspects (Baek et al., 2011; Dhir et al., 2015; Karnik et al., 2013; Smock et al., 2011). Understanding the feature-specific aspects of a particular SNS is considered significant as in recent years, the number of features (e.g. chatting, photo tagging, photo sharing, wall postings, etc.) have increased considerably on SNS, and notably Facebook. These features offer varying activities, and the relevant motivations and usages differ substantially among the SNS users (Joinson, 2008; Smock et al., 2011). By embracing these endorsements, this dissertation contributes to the emergent literature exploring feature-specific aspects of Facebook (Cheng & Leung, 2015; Dhir et al., 2015; Krause et al., 2014).

Third, the dissertation work also contribute towards expanding the U&G theory to some extent. For example, results from Study III revealed "seeking affection" (in the form of likes and comments) as one of the six identified U&G of photo sharing on Facebook. This U&G is a novel one that has not been identified by prior literature on U&G of specific SNS features.

Fourth, by examining the Facebook photo tagging practices among Indian adolescents, this dissertation contributes to the limited literature on SNS practices by adolescents in the developing world (Study II). Prior research on SNS behaviors has predominantly focused on university students and adults in the developed countries, notably the USA (Park & Lee, 2014; Sheldon & Bryant, 2016). Consequently, study II of this dissertation is one of the very few studies that examines the behavior of adolescent SNS users in developing countries. It is extremely important to understand adolescent Facebook users since they are active SNS users who are the least-studied in the prior literature. Furthermore, studying SNS users from India and developing countries is important since they are quickly emerging as the dominant SNS users. In a broader context, this dissertation also contributes theoretical knowledge to the domain of ICT for development (ICT4D).

Fifth, the results from this dissertation contribute to a number of multidisciplinary research domains from the perspective of privacy and security, as it provides insights into various aspects of privacy associated specifically with photo sharing on Facebook (Study IV, V). Based on the extensive literature on privacy in various computer-mediated platforms and SNS, study IV developed and tested a model to validate some of the key privacy variables that can influence trust and users' activity, and consequently photo sharing on Facebook. The results of this empirical model add to the need emphasized by scholars for empirical research that accommodates multidimensional constructs of privacy, trust, and user intentions to share content (Dwyer et al., 2007; Shin, 2010). The model proposed in study IV can be useful for researchers interested in understanding the relationship of various aspects of privacy with trust, user activity, and users' intentions to various forms disclosures on Facebook such as status updates, location check-ins, and instant messaging. Furthermore, replicating this model to understand the relationship of privacy and trust on specific SNS features such as video sharing, group participation, and interoperability could also provide unique insights.

## 4.2   Practical implications

This dissertation has concluded with various practical implications for practitioners and scholars engaged in the field of SNS, new media research, content/application design, as well as companies promoting their products and services on SNS. Furthermore, some practical implications can also apply to governmental organizations, consumer rights agencies, and the general public.

First, in general, understanding how and why users engage with digital photos will be helpful, as in recent years popularity and adoption of photos have increased substantially on various SNS. Considering the identified uses and gratifications of digital photos, companies engaged in promoting their products and services on SNS platforms can realign their visual as well as other forms of content strategies accordingly.

Second, by considering uses and gratifications of digital photos highlighted in this dissertation, companies can produce digital photos that attract their users and offer a higher possibility to be re-shared by them. For instance, the results from study III indicate that users are motivated to share photos to fulfill attention seeking needs, therefore rewarding the users in the form of virtual gifts can further facilitate the digital-photo interaction on brand pages and groups. These virtual rewards can potentially lead to increased devotion from the existing users as well as a broader customer base. Third, the designers and developers engaged in framing new concepts and solutions for various computer-mediated platforms that utilize photos or other multimedia content could benefit from the study results. Fourth, media companies and campaigners can also utilize these findings to extend the reach of their digital content on SNS platforms.

Fifth, the findings of this dissertation on demographic attributes such as age and gender differences can also be helpful for companies targeting specific user demographics. These findings can also provide insight to scholars interested in understanding age and gender differences in computer-mediated platforms. Sixth, the results from this dissertation can also be valuable for SNS providers, media companies, and designers interested in various aspects of privacy and trust. The results from study IV and V (and to some extent study II) provide valuable insights into these aspects and how these elements play a role in encouraging or limiting digital photos usage and adoption on Facebook. The model proposed in study IV can also be replicated to other Facebook features. This approach can be helpful in grasping the complex concept of SNS privacy by studying the concept at a very basic level. Moreover, features and applications on other SNS platforms can also utilize the proposed model to relate privacy aspects with trust and users' intentions for that specific feature/application. Even though, results from study IV validate the findings from prior research that privacy concerns do not significantly impact users' behaviors (boyd & Hargittai, 2010; Debatin et al., 2009), they are very high across the studied respondent groups. SNS platforms should address these concerns by ensuring their users that the content shared by them is well-protected. This assurance will encourage and motivate users towards more frequent content interaction on these platforms in the future. The SNS users should also be educated not only about the possible privacy and security threats on SNS and how to mitigate them but also about respecting the privacy of others and their content in computer-mediated platforms. Likewise, governmental agencies should devise policies to alleviate threats posed not only by organizations but from the society as well, as the majority of users seem more concerned about social threats than the organizational threats. Results from this dissertation indicating privacy awareness is somewhat at a moderate level also call for relevant stakeholders to provide awareness to SNS users about online data privacy. Considering the wide diversity of SNS users across the globe, SNS should put more effort into simplifying their privacy policies and privacy settings so they are more comprehensible. For instance, to ensure that the content of a newly-joined user is well-protected, all the privacy settings should be set at the most restrictive level by default.

## 4.3   Study limitations and recommendations for future work

Despite the number of contributions, this dissertation work also has a few limitations that are briefly discussed in this section. These limitations offer new opportunities for future research within multiple domains. Furthermore, accommodating these limitations will be beneficial to reproduce more reliable and generalizable study results in the future. First, the generalizability of the study results can be considered to be one of the main limitation of this dissertation work. Study III, IV, and V were based on online surveys that were only promoted on a limited number of randomly yet cautiously selected Facebook groups. Even though we made all possible efforts to get a representative sample, due to our limited reach and timeframe, the convenience sample is likely to be non-representative. Due to this reason, the findings might not be generalized to all the Facebook users. To minimize this bias, future studies should avoid focusing solely on a single channel. Instead, we would recommend employing a multitude of channels to promote online surveys. Furthermore, the age categories used in both the surveys are different, hence, it is difficult to compare both the surveys from the perspective of age groups.

Second, due to a number of constraints, (including data gathering challenges and certain requirements of the funded projects), these studies were carried out with diverse ranging populations that also limit the generalizability of the study results. The results from the study could have been more concrete, thorough and applicable if a particular set of users would have been studied.

Third, even though the number of study samples in both the surveys are comparable to similar kind of studies (Hollenbaugh & Ferris, 2014; Park & Lee, 2014), the respondents from these surveys were predominantly from developed countries including the USA and European countries. Hence, these findings might also not provide a representative picture of Facebook users due to the limited number of respondents from developing countries. It is important to highlight that Facebook and other SNS are increasingly popular among users from developing world, for instance; Facebook  is currently ranked in the top 3 websites of most of these countries ("Alexa Top 500 Global Sites," 2015). Hence, carrying out similar studies with Facebook users in developing countries can be helpful in improving the generalizability and validity of the results. Furthermore, replicating similar studies in developing countries could potentially lead to the development of novel and interesting insights that might be pertinent to those regions only.

Fourth, the online surveys also lacked an extensive set of demographic variables. Furthermore, as these surveys were not targeted towards a specific group of user demographics, it is possible that the results cannot apply to a specific group of users. Fifth, as most of the results from this dissertation are specifically tied to digital-photo interaction on Facebook, they might not be applied directly to other SNS platforms such as Twitter, Snapchat, or Instagram. Future studies can consider studying gratifications and various aspects of privacy related to digital-photo interaction specific to other popular SNS platforms.

Sixth, the survey on privacy aspects of digital photo sharing employed a limited number of questions to measure various aspects of privacy, and no clear distinction were made between social and organizational privacy concerns. Future work can extend further by adding new dimensions of privacy with a clear distinction between social and organizational concerns of privacy. Future work can also link the gratifications and privacy aspects with other factors commonly highlighted in the literature including SNS usage frequency, time spent on the platform, and personality attributes of SNS users, e.g. the five big factors. Moreover, study V relied solely on bivariate analysis for identifying age and gender differences. Furthermore, employing regression analysis to the survey results could have also identified interesting relationships between the investigated variable. Finally, the results of Study II also have limited generalizability, as it was carried out with Indian adolescents enrolled in schools where the medium of instruction is English. These results are confined to a very specific group of teenagers in India, and the study results might not even apply to teenagers across India or other developing countries.

## 4.4 Conclusion

Motivated by the enormous popularity of digital photos on numerous computer-mediated platforms, this dissertation work provides a multidisciplinary perspective on various fields of research including computer-mediated communication, human-computer interaction, SNS behaviors, and new media. Despite having some aforementioned limitations, this dissertation extends our current understanding of how and why users on computer-mediated platforms engage and interact with digital photos. Furthermore, how various factors interrelate and impact these interactions on these platforms is also addressed. This dissertation work contributes notably in further extending the current literature on feature-specific U&G of SNS platforms. Moreover, by providing a granular understanding of privacy and trust relationship with digital photos, this dissertation has further progressed the existing knowledge pertaining to privacy in SNS literature. This work has also investigated the relationship of digital photos with age and gender which are considered as some of the crucial determinates of SNS usage. Finally, the work also contributes to the literature as well as our current understanding of ICT and SNS usage by adolescents in the developing world.

To achieve the above-stated contributions, this dissertation work utilized qualitative and quantitative research methods for collecting data including online surveys, open-ended questionnaire, and task evaluation activities. This dissertation work achieved the desired research objectives through five empirical articles that supplement the framing of this dissertation. The findings of this dissertation work will not only benefit academic scholars and practitioners in various domains, but also SNS companies, governmental and consumer rights agencies, and society at large. Finally, extending this work as proposed in the previous section will significantly extend our current understanding to factors that are relevant to digital-photo interaction in computer-mediated platforms.

The last decade has witnessed an ever-increasing usage and popularity of digital photos on numerous computer-mediated platforms. However, the understanding of the motivations and behaviors associated with photos related practices and factors that influence these practices remain largely unmapped. This dissertation aims to explore some of the pertinent issues that shape "digital-photo interaction", i.e. user practices, motivations, and tensions that exist among various photos related features and activities. To bridge the research gap, this dissertation primarily concentrates on two highly popular activities i.e. photo-sharing and photo-tagging.

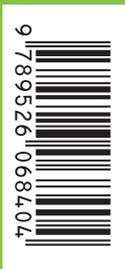

BUSINESS +
ECONOMY

ART +
DESIGN +
ARCHITECTURE

SCIENCE +
TECHNOLOGY

CROSSOVER

DOCTORAL
DISSERTATIONS